%% file: main.tex
\documentclass[lettersize,journal]{IEEEtran}
\IEEEoverridecommandlockouts
\usepackage{subcaption}
\usepackage[ruled,vlined,linesnumbered]{algorithm2e}
\usepackage{xspace}
\usepackage{cite}
\usepackage{amsmath,amssymb,amsfonts}
\usepackage{algorithmic}
\usepackage{graphicx}
\usepackage{textcomp}
\usepackage{xcolor}
\usepackage{booktabs}
\usepackage{multirow}
\usepackage{caption}
\newcommand{\xxx}{DeepFeature\xspace}

\newcommand{\fma}{FMA\xspace}

\usepackage[figuresright]{rotating}
\newcommand{\fvs}{FVS\xspace}
\newcommand{\fas}{FAS\xspace}
\usepackage[capitalize]{cleveref}
\crefname{section}{Sec.}{Secs.}
\Crefname{section}{Section}{Sections}
\Crefname{table}{Table}{Tables}
\crefname{table}{Tab.}{Tabs.}
\usepackage[numbers]{natbib}

\newcommand{\ie}{i.e., }

\usepackage{tcolorbox}
\newcommand{\mybox}[1]{\begin{tcolorbox}[colback=yellow!10!white,colframe=red!75!black,lowerbox=invisible,savelowerto=\jobname_ex.tex]\emph{#1}\end{tcolorbox}}

\begin{document}

% \author{
% 	\IEEEauthorblockN{
% 		Dong Huang\textsuperscript{1}\textsuperscript{$\dagger$}\thanks{\textsuperscript{$\dagger$}These authors contributed equally to this work}, 
% 		Qingwen Bu\textsuperscript{2}\textsuperscript{3}\textsuperscript{$\dagger$}, 
%         Yichao Fu\textsuperscript{1}, 
% 		Yuhao Qing\textsuperscript{1}, 
% 		Bocheng Xiao\textsuperscript{1} 
% 		Heming Cui\textsuperscript{1}\textsuperscript{3}
% 	\IEEEauthorblockA{\textsuperscript{1}The University of Hong Kong}
% 	\IEEEauthorblockA{\textsuperscript{2}Shanghai Jiao Tong University}
%    \IEEEauthorblockA{\textsuperscript{3}Shanghai Artificial Intelligence Laboratory}
%     \IEEEauthorblockA{\{dhuang, yhqing, heming\}@cs.hku.hk, qwbu01@sjtu.edu.cn, yichao@connect.hku.hk}
%  }
% } 
\author{Dong Huang\textsuperscript{1}\textsuperscript{$\dagger$}\thanks{\textsuperscript{$\dagger$}These authors contributed equally to this work.}, 
  Qingwen Bu\textsuperscript{2}\textsuperscript{3}\textsuperscript{$\dagger$}, 
  Yuhao Qing\textsuperscript{1}, 
  Yichao Fu\textsuperscript{1},
  Heming Cui\textsuperscript{1}\textsuperscript{3}\\
\textsuperscript{1}The University of Hong Kong\\
\textsuperscript{2}Shanghai Jiao Tong University\\
\textsuperscript{3}Shanghai Artificial Intelligence Laboratory\\
\{dhuang,yhqing,heming\}@cs.hku.hk,qwbu01@sjtu.edu.cn,yichao@connect.hku.hk}

\title{Feature Map Testing for Deep Neural Networks}

\maketitle

\begin{abstract}

Due to the widespread application of deep neural networks~(DNNs) in safety-critical tasks, deep learning testing has drawn increasing attention. During the testing process, test cases that have been fuzzed or selected using test metrics are fed into the model to find fault-inducing test units (e.g., neurons and feature maps, activating which will almost certainly result in a model error) and report them to the DNN developer, who subsequently repair them~(e.g., retraining the model with test cases). Current test metrics, however, are primarily concerned with the neurons, which means that test cases that are discovered either by guided fuzzing or selection with these metrics focus on detecting fault-inducing neurons while failing to detect fault-inducing feature maps.

In this work, we propose \xxx, which tests DNNs from the feature map level. When testing is conducted, \xxx will scrutinize every internal feature map in the model and identify vulnerabilities that can be enhanced through repairing to increase the model's overall performance. Exhaustive experiments are conducted to demonstrate that (1) \xxx is a strong tool for detecting the model's vulnerable feature maps; (2) \xxx's test case selection has a high fault detection rate and can detect more types of faults~(comparing DeepFeature to coverage-guided selection techniques, the fault detection rate is increased by 49.32\%). (3) \xxx's fuzzer also outperforms current fuzzing techniques and generates valuable test cases more efficiently.
\end{abstract}


% \footnotetext{These authors contributed equally to this work.}
\input{intro}
\input{background}
\input{methodology}

\input{evaluation}
\input{related}

\input{conclusion}

\bibliographystyle{IEEEtranN}
\bibliography{IEEEfull}

\end{document}

%% file: intro.tex
\section{Introduction}

Deep neural networks (DNNs) are widely used in safety-critical domains, such as autonomous driving~\cite{Bojarski2016EndTE} and medical diagnosis~\cite{Rajpurkar2017CheXNetRP}. However, DNNs can be complex and susceptible to data bias, overfitting, and underfitting, making them far from dependable for these applications. To address this problem, DNNs must undergo deep learning testing techniques before deployment to ensure their reliability~\cite{pei2017deepxplore, ma2018deepgauge}. This testing helps identify and repair units~(e.g., neurons and feature maps) in the model that are particularly vulnerable to change and can significantly impact the model's performance, resulting in a more reliable model.

The deep learning testing technique mainly has three key elements: coverage metric, test case selection, and fuzzing strategy. The testing process typically works in an iterative way. In each iteration, the testing techniques will first feed test cases, which are selected or fuzzed with certain rules~(e.g., coverage metric), into the model, after which a coverage metric is calculated. New test cases will be generated to further improve the coverage metric until the metric meets the developers' requirements.

\begin{figure}[t]
    \centering
    \includegraphics[width=\linewidth]{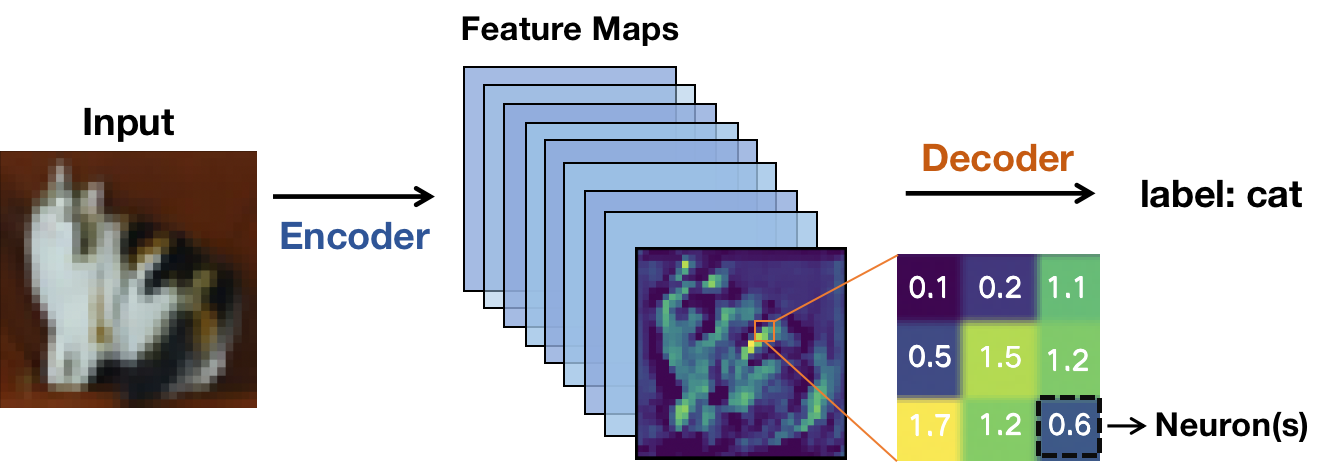}
    \caption{An illustration of feature maps and neurons. 
    A neuron is a basic unit of computation in a neural network. A feature map is a set of neurons that are connected to the same input and that together produce an output that encodes certain features~(e.g., color, and texture).}
    \label{fig:vis}
    % \vspace{-0.7cm}

\end{figure}

The most important element mentioned above is the coverage metric since it determines how test cases are generated~(i.e., fuzzed or selected from a massive amount of candidate test sets) to reveal the fault-inducing units~(e.g., neuron and feature map. Illustrated in~\cref{fig:vis}) of a tested DNN. Therefore, the main criterion of coverage metrics is to explore as much diversity as possible of a certain subspace defined based on different abstraction levels~(e.g., neuron activation~\cite{pei2017deepxplore}, neuron activation pattern~\cite{ma2018deepgauge}, and neuron activation conditions~\cite{Sun2018ConcolicTF}), while test case selection and fuzzing strategy are used to obtain new test cases, which are fed into the model to increase coverage metrics. To explore more types of model units' behaviors, multiple \textbf{N}euron \textbf{C}overage~(NC) metrics, including NAC~\cite{pei2017deepxplore}, KMNC~\cite{ma2018deepgauge}, IDC~\cite{Gerasimou2020ImportanceDrivenDL}, and MC/DC~\cite{Ma2018DeepMutationMT} have been proposed.

\noindent\textbf{Key problems of existing NC-guided testing.}
Despite the fact that existing NC-guided testing techniques have greatly advanced deep learning testing, there remains a significant constraint, namely that not all fault-inducing units in the model are neurons. Recent studies~\cite{jahangir2022FMadversarial,wang2021feature} have revealed that feature maps are also important fault-inducing units that can affect DNN performance, while NC-guided techniques often focus on analyzing neuron's behavior and fail to capture and utilize feature-map-level information, causing DNN developers can not detecting feature maps that induce errors. Consequently, NC-guided testing techniques yield a sub-optimal fault detection rate and can merely bring a significant boost to the model~\cite{ismeanful,Structural}, as retraining the model with test cases generated by NC-Fuzz~(e.g., DeepXplore~\cite{pei2017deepxplore}, DeepHunter~\cite{Xie2018DeepHunterHD}, and ADAPT~\cite{ADAPT}) can only repair fault-inducing neurons.

In this work, we address this problem by proposing \xxx, which test DNNs from feature map level. 
As opposed to heuristically defining various neuron-level metrics, we establish a clear connection between output variation at the feature-map level and the performance of the model through clear experiments.
As illustrated in~\cref{fig:vis}, the main distinction between \xxx and existing neuron coverage testing works is that \xxx tests the DNN model from the feature-map level. 
Specifically, during the testing process, \xxx delves into every inner feature map in the model to evaluate its vulnerability.
The vulnerable feature maps therein are then repaired by retraining with test cases selected or generated using our corresponding selection and fuzzing strategy while further improving the overall performance of the model.

To validate the effectiveness of \xxx, we conduct experiments with four popular DNN models across four widely-used datasets. The experiment results validate our motivation and demonstrate the superiority of our method.
Specifically, in comparison to state-of-the-art neuron-level test case selection methods, our \xxx yields a higher fault detection rate~(e.g., DeepFeature maximum increases 49.32\% fault detection rate compared with NC-guided baselines), and in limited test case selection scenarios, \xxx can detect more type of faults that are ignored by baselines.
\xxx's Fuzzer is further exhibited to be the most effective and efficient among state-of-the-art fuzzing techniques.

In a nutshell, we make the following contributions.
\begin{itemize}
    \item We propose a novel set of testing metrics to quantify the feature map's vulnerability.
    
    \item We propose a new selection method that can productively select test cases with a high probability of fooling the model from a massive number of unlabeled test cases~(e.g., DeepFeature maximum increases 49.32\% fault detection rate compared with NC-guided baselines).
    
    \item We propose feature-map-guided fuzzing technique that outperforms current fuzzing techniques.
\end{itemize}

%% file: background.tex
\section{Background}

This section includes several basic knowledge of neural networks, neuron coverage metrics, test case selection, and fuzzing strategies.

\subsection{Neuron, Feature Map and Neural Network}
In this work, we focus on deep learning models~(e.g., deep neural networks) for classification, which can be presented as a complicated function $f: \mathcal{X} \rightarrow \mathcal{Y}$ mapping an input $x \in \mathcal{X}$ into a label $y\in\mathcal{Y}$. Unlike traditional software, programmed with deterministic algorithms by developers, DNNs are defined by the training data, along with the network structures. 
% Generally speaking, a DNN consists of an input layer, an output layer, and at least one hidden layer. Each neuron in each layer is intertwined with neurons in other layers, and the output of each neuron is the weighted sum of the outputs of all neurons in the previous layer and then a nonlinear activation function (e.g.,tanh, sigmoid, and ReLU) is applied.

% % \paragraph{Feature Map and Neuron}
% \subsection{Neurons and Feature Maps in DNNs}

As shown in~\cref{fig:vis}, neurons and feature maps are fundamental concepts in the field of deep neural networks (DNNs), which have emerged as powerful tools for image recognition and other tasks involving large amounts of data.
A neuron is a basic unit of computation in a neural network. It receives input from other neurons or from the input layer, applies a set of weights and biases to this input, and then applies an activation function to produce an output, where the activation function is a mathematical function that determines the output of a neuron based on its input. Common activation functions include Sigmoid, Tanh, and ReLU.
A feature map, on the other hand, is a set of neurons that are connected to the same input and that together produce an output that encodes certain features or patterns in the input. In a DNN, multiple feature maps are typically stacked together to form a hierarchy of features, with lower-level feature maps encoding simpler patterns and higher-level feature maps encoding more complex patterns.

The main difference between neurons and feature maps is that neurons are individual units of computation, whereas feature maps are collections of neurons that work together to extract specific features from the input. Neurons are the building blocks of a DNN, whereas feature maps are the building blocks of the DNN's ability to recognize patterns in the input data.

\subsection{Neuron Coverage Metrics}

\paragraph{\textbf{Neuron Activation Coverage~(NAC($k$))}}
NAC($k$) was proposed by DeepXplore~\cite{pei2017deepxplore}, NAC($k$) assumes that the more neurons are activated, the more states of DNN are explored. The parameter $k$ of this coverage criteria is defined by the developer to specify how a neuron in a DNN can be counted as covered~(i.e., if the output of a neuron is large than $k$, the developer will take the neuron as covered). The rate of NAC($k$) for a test is defined as the ratio of the number of covered neurons to the total number of neurons. 

\paragraph{\textbf{K-multisection Neuron Coverage~(KMNC($k$))}}
Based on the NAC($k$) assumption about the DNN states, DeepGuage~\cite{ma2018deepgauge} further partitions the neuron's output into $k$ ranges~(e.g., 2000), and each range represents one state in DNN. Specifically, suppose that the output of a neuron $N$ on the training set is in the interval $[low_{N}, high_{N}]$, and divide them equally into $k$ segments. 
% The goal of the KMNC($k$) criteria is to make the neuron cover each segment of $k$ segments.

% \paragraph{\textbf{Neuron Boundary Coverage~(NBC)}}
% Different from KMNC($k$), NBC does not aim to cover all sections in $[low_{N}, high_{N}]$, it focuses on whether the corner regions $(- \infty, low_{N}]$ and $[high_{N},\infty)$ are covered by test cases~\cite{ma2018deepgauge}. The rate of NBC is defined as the ratio of the number of covered boundaries to the total number of boundaries. Since each neuron has one upper bound and one lower bound, the total number of boundaries is twice the number of neurons.

% \paragraph{\textbf{Strong Neuron Activation Coverage~(SNAC)}}
% SNAC measures how many corner cases have been covered by the given test cases~\cite{ma2018deepgauge}. Specifically, SNAC can be regarded as a special case of NBC as it only takes upper boundaries $[high_{N},\infty)$ into consideration. Thus, it is defined as the ratio of the number of covered upper boundaries to the total number of upper boundaries, in which the latter is equal to the number of neurons in a DNN.

% \paragraph{\textbf{Top-k Neuron Coverage~(TKNC($k$))}}
% TKNC($k$) focuses on the most active $k$ neurons in each layer~\cite{ma2018deepgauge}. It is defined as the ratio of the total number of top-$k$ neurons in each layer and the total number of neurons in a DNN.

\subsection{Test Case Selection Methods}
In this section, we introduce several widely used test case selection methods, which are used to select valuable test cases from a massive number of datasets. Specifically, the goal of the test case selection method is to sampler a fixed size~($N$) subset $I_{N}$ from the total test set $I_{T}$. We divide test case selection methods into two types: coverage-guided test case selection and prioritization test case selection.
\subsubsection{Coverage-guided test case selection}
Coverage-guided~(e.g., NAC) selection methods try to select test cases that can reach maximum coverage metrics and lead to a higher fault detection rate~\cite{pei2017deepxplore,kim2019guiding,ma2018deepgauge}.

\subsubsection{Prioritization test case selection}
Generally, for a given total test set $I_{T}$, prioritization test selection methods compute a probability $p_{i}$ for each test case $i$ in the total test set. The value of $p_{i}$ represents the probability of test case $i$ sampled by selection methods.
% Here we introduce one popular prioritization-guided test case selection.
% \paragraph{\textbf{DeepGini}}
% \citet{deepgini} propose a test prioritization technique based on a statistical perspective of DNN, named DeepGini. It takes the use of the Gini coefficient to measure the likelihood of test case $i$ being misclassified.

\subsection{Fuzzing Strategies}
DeepXplore~\cite{pei2017deepxplore} is the first DNN testing framework. It proposes a neuron activation coverage~(NAC) guided fuzzing strategy to increase the neuron activation coverage metric. Inspired by DeepXplore, ~\citet{DeepMutation++} propose DeepMutation++~\cite{DeepMutation++}, which generates test cases to increase multi-granularity coverage metrics.
Inspired by DeepXplore, and DeepMutation++, ~\citet{ADAPT} propose ADAPT, which integrates multiple neuron behaviors ~(e.g., NAC and TKNC) to fuzz test cases.

\begin{table}[]
    % \small
    % \begin{centering}
    % \setlength{\tabcolsep}{2.5pt}
    \centering
    \begin{tabular}{ccccc}
    \toprule
    Dataset&DNN Model &Neurons & Layers&Ori Acc (\%)\\
    \midrule
    \multirow{2}*{MNIST~\cite{deng2012mnist}} & LeNet-1~\cite{lecun1998gradient}&3,350&5&89.50\\
    ~&LeNet-5~\cite{lecun1998gradient}&44,426&7&91.79\\
    \midrule
    \multirow{2}*{CIFAR-10~\cite{Krizhevsky09learningmultiple}}& ResNet-20~\cite{he2016deep}&543,754&20&86.07\\
    ~&VGG-16~\cite{simonyan2014very}&35,749,834&21&82.52\\
    \midrule
    \multirow{2}*{Fashion~\cite{xiao2017/online}}& LeNet-1~\cite{lecun1998gradient}&3,350&5&78.99\\
    ~&ResNet-20~\cite{he2016deep}&543,754&20&86.12\\
    \midrule
    \multirow{2}*{SVHN~\cite{Netzer2011}} & LeNet-5~\cite{lecun1998gradient}&44,426&7&84.17\\
    ~&VGG-16~\cite{simonyan2014very}&35,749,834&21&92.02\\
    \bottomrule
    \end{tabular}
    \caption{Datasets and DNNs for evaluating \xxx.}
    % \vspace{-0.8cm}
    \label{tab:dataset}
\end{table}

%% file: methodology.tex
\section{Methodology}

\begin{figure}[t]
    \centering
    \includegraphics[width=1\linewidth]{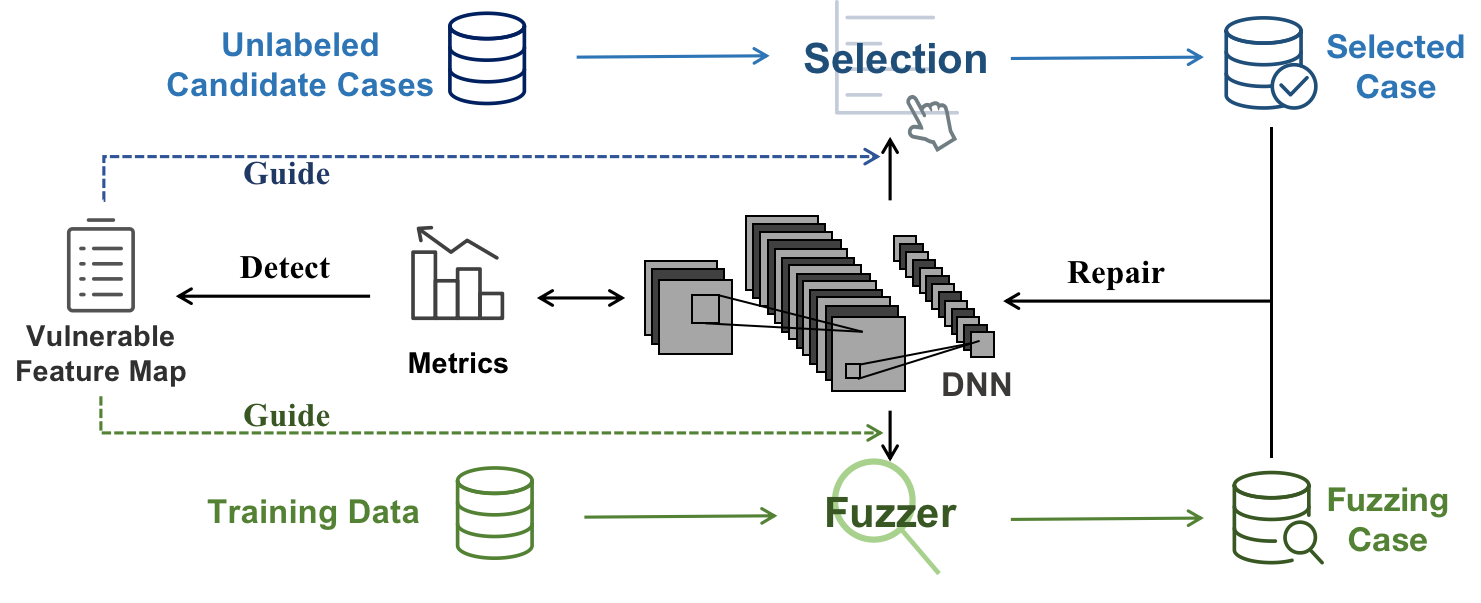}
    \caption{Overview of \xxx's framework.}
    \label{fig:pipeline}
    % \vspace{-0.4cm}
\end{figure}
\subsection{Motivation}

Deep Neural Networks (DNNs) are known to be sensitive to any perturbations in their parameters, leading to vulnerabilities in certain units~(e.g., neurons or feature maps), making them more susceptible to perturbations~\cite{Wang2021RobOTRT,pei2017deepxplore,Xie2018DeepHunterHD}. Deep learning testing aims to discover these vulnerable units and repair them to make the model more reliable. Traditional deep learning testing has only considered the vulnerability of neurons by generating test cases to vary certain neurons' status and see if a DNN produces wrong labels, but recent studies~\cite{wang2021feature,jahangir2022FMadversarial} have shown that without altering any neuron's status, feature maps (sets of neurons) can also lead to wrong DNN output and thus become vulnerable. Therefore, a feature map may not contain any vulnerable neurons but can still be vulnerable and cause a significant impact on the model's performance. 

Unfortunately, current neuron testing techniques are not equipped to detect vulnerable feature maps due to a lack of proper metrics, evaluation methods for vulnerable feature maps, and testing techniques designed to detect vulnerable feature maps. Existing coverage metrics for deep learning testing are based on neurons, and there are no methods to evaluate the vulnerability of a feature map. As a result, there is a pressing need for a comprehensive testing approach that addresses these limitations.

In this section, we address above-mentioned problem by proposing \xxx, a novel feature-map-level testing technique for deep neural networks.
\xxx can detect vulnerable feature maps in a DNN and repair them.
The overall framework of \xxx is shown in~\cref{fig:pipeline}.
For ease of discussion, this section defines the following notations for DNNs and feature maps:
$f_\theta$ is a DNN parameterized by $\theta$, and there are $C$ channels (\ie $C$ different feature maps) in the model. The $c$-th feature map in the DNN is denoted as $f_{\theta}^{c}$. $f_{\theta}^{c}(x)$ denotes the output of the corresponding feature map when the network input is $x$.

\subsection{Feature Map Attack}
We first propose the prototype of Feature Map Attack~(FMA), which is used to generate mutation cases for a given model's feature map and test cases. The FMA algorithm, shown in~\cref{algo:attack}, iteratively generates mutated samples and maximizes the difference between the feature maps of the mutated cases and the original test cases. Specifically, the algorithm first initializes each test case with added random noise and then performs a specified number of iterations, using a gradient-based optimization method to maximize the difference between the feature maps of the mutated cases and the original test cases. In our algorithm, we add a clamping function to ensure that the perturbations stay within a certain range. We also set the number of iterations as $steps$ while the parameter $\epsilon$ controls the strength of the perturbations. 

% \subsection{Feature map Testing Metrics}\label{sec:metrics}

% \begin{algorithm}
%     \caption{Feature Map Attack}\label{algo:attack}
%         \SetKwInOut{KOutput}{output}
%     \SetKwProg{Fn}{Function}{:}{}
%     \SetKwFunction{Fun}{Fun}
%     \KwIn{
%     $f_{\theta}^c$: model's feature map to be tested;
%     $x$: test case~(clean sample);
%     $\alpha$: step size;
%     $\epsilon$: maximum perturbation;
%     $steps$: mutation steps;
%     }
    
%     \KOutput{
%     $x'$ mutation case~(adversarial sample).
%     }

%          \Fn{FMA$(f_{\theta}^c, x, steps,\alpha,  \epsilon)$}{

%         Initialize: $x^{'} \leftarrow x + random\_noise$\;
%     %  remain\_steps \leftarrow steps\;
%      \For{$step=0$ to $steps$}
%      {
%           $loss$  \leftarrow\  \Vert$f_{\theta}^{c}(x) -  f_{\theta}^{c}(x')\Vert ^2$\;
          
%           grad \leftarrow \ $\partial {loss}\ /\  \partial x' $\;
%           $x^{'}$ \leftarrow\  $x^{'} + \alpha * $\ grad.sign()\;
%           $\delta$ \leftarrow clamp($x^{'}-x$, $ min=-\epsilon$, $max=\epsilon$)\;
%           $ x^{'} $ \leftarrow clamp($x\ +\ \delta$,$min=0$,$max=1$)\;
%         %   $remain\_steps$ \leftarrow remain\_steps - 1\;
%      }

%     \Return $x'$
%     }
% \end{algorithm}

\begin{algorithm}
    \caption{Feature Map Attack}\label{algo:attack}
    \SetKwInOut{KOutput}{output}
    \SetKwProg{Fn}{Function}{:}{}
    \SetKwFunction{Fun}{Fun}
    
    \KwIn{
    $f_{\theta}^c$: model's feature map to be tested;
    $x$: test case (clean sample);
    $\alpha$: step size;
    $\epsilon$: maximum perturbation;
    $steps$: mutation steps;
    }
    
    \KOutput{
    $x'$: mutation case (adversarial sample).
    }

    \Fn{$\text{FMA}(f_{\theta}^c, x, steps, \alpha, \epsilon)$}{

        Initialize: $x^{'} \leftarrow x + \text{random\_noise}$\;
     
     \For{$step=1$ $\KwTo$ $steps$}{
          $loss \leftarrow \Vert f_{\theta}^{c}(x) - f_{\theta}^{c}(x')\Vert^2$\;
          
          $grad \leftarrow \frac{\partial loss}{\partial x'}$\;
          
          $x^{'} \leftarrow x^{'} + \alpha \cdot \text{sign}(grad)$\;
          
          $\delta \leftarrow \text{clamp}(x^{'}-x, - \epsilon, \epsilon)$\;
          
          $x^{'} \leftarrow \text{clamp}(x + \delta, 0, 1)$\;
     }

    \Return $x'$
    }
\end{algorithm}

\subsection{\textbf{F}eature map \textbf{A}ttack \textbf{S}core}\label{sec:fac}
The attack success rate of DNNs under FMA against different feature maps can vary greatly. Generally, different feature maps in a DNN have different vulnerabilities, and their impact on model attack success rate is also different. Therefore, we propose Feature map Attack Score~(\fas) to measure the effect of different feature maps on model performance. 
The \fas for the $c$-th feature map is defined as the attack success rate of the DNN when the $c$-th feature map is attacked by \fma. 
Specifically for the $c$-th feature map in the model. $\fas(f_{\theta}^{c}, \mathcal{X})$ is defined as:

$$\fas(f_{\theta}^{c}) = 1 - \frac{\sum_{i=0}^{N}\mathbb{I}(f_{\theta}(x_i')==y_i)}{\sum_{i=0}^{N}\mathbb{I}(f_{\theta}(x_i)==y_i)}$$

Where $\mathcal{X} = \{x_i \mid i = 1, 2, \dots, N\}$ is the developer provided test cases, $\{y_i \mid i = 1, 2, \dots, N\}$ is its ground-truth label and the mutation samples $\{x_i' \mid i = 1, 2, \dots, N\}$ is generated by \fma based on the feature map $f_{\theta}^{c}$~(i.e., for different feature map, \fma will generate different mutation samples).
The \fas for a feature map ranges between 0 and 1, where the value of 0 means that using \fma to attack the feature map will not affect the performance of the DNN, which indicates the feature map is not vulnerable.
The value of 1 means that using \fma to attack the feature map has a significant effect on the performance of the DNN, and the larger the \fas value means the feature map is more vulnerable than others.

\begin{table}[t]
    \centering
    \small
    
    \setlength{\tabcolsep}{3mm}{
    \begin{tabular}{c|cccc}
    \toprule
         Feature Map &\multicolumn{4}{c}{\fma steps}  \\
         Index&3&7&10&20\\
         \midrule
         1&58.85&68.73&68.65&67.72\\
         2&35.15&44.09&44.11&43.07\\
         3&3.45&3.72&3.70&3.91\\
         4&48.26&60.98& 59.29& 62.81\\
         5&32.90&42.42&42.74&41.09\\
         6&56.68&66.07&66.93&66.39\\
         \bottomrule
    \end{tabular}}
    \caption{A example of FAS in LeNet5 pretrained model}
    \label{tab:example_fac}
\end{table}
\begin{figure}[t]
% \tiny
    \centering
    \begin{subfigure}[b]{0.44\linewidth}
        \subcaption[short for lof]{SVHN\&LeNet-5}
        \includegraphics[width=\linewidth]{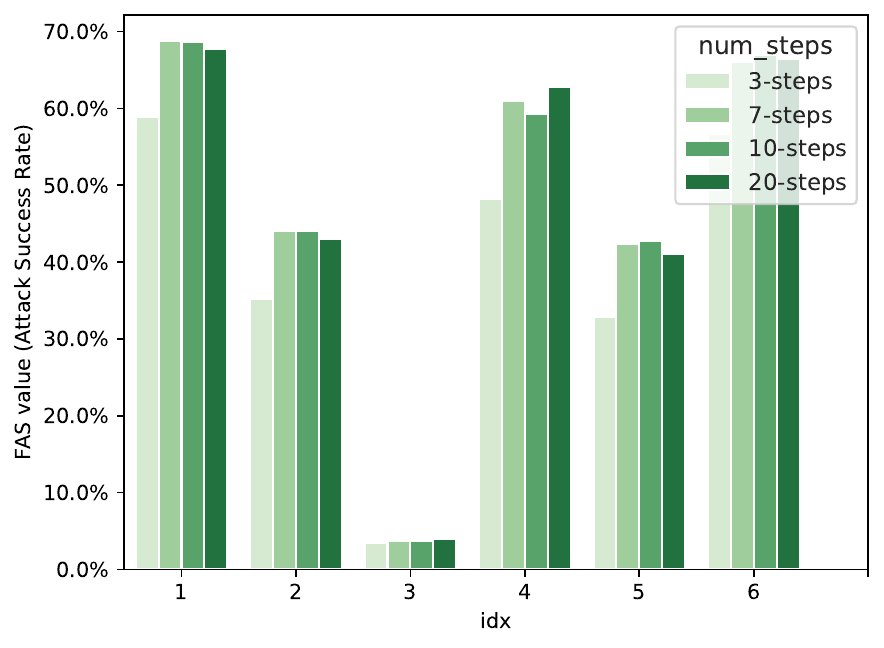}
        \label{fig:subfig1-1}
    \end{subfigure}
     \begin{subfigure}[b]{0.44\linewidth}
        \subcaption[short for lof]{CIFAR10\&ResNet-20}
        \includegraphics[width=\linewidth]{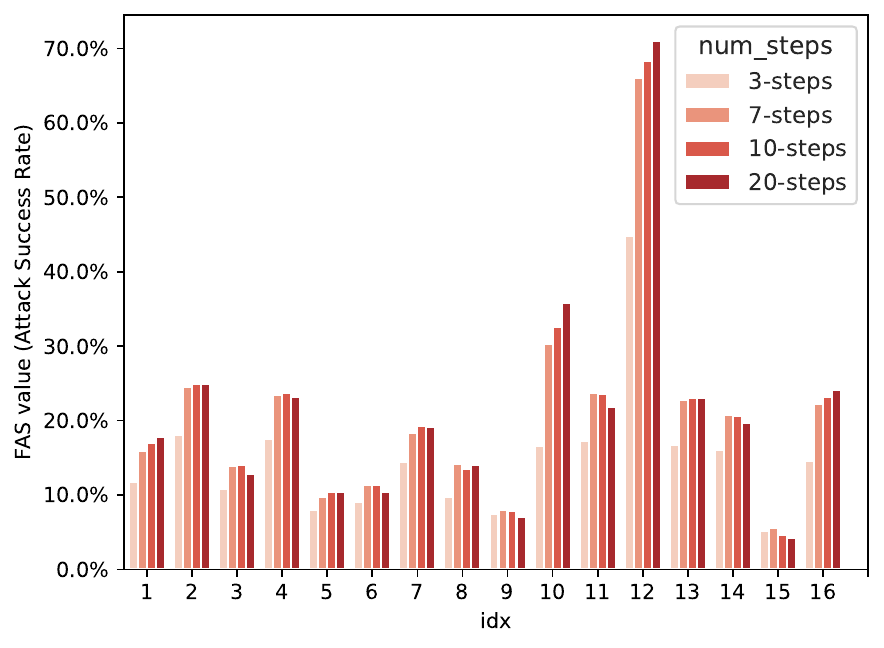}
        \label{fig:subfig1-2}
    \end{subfigure}%
    \caption{Different feature maps' vulnerability. The x-axis is the index of the feature maps, and the y-axis is the \fas value~(attack success rate). A larger \fas value indicates a more vulnerable feature map.}
        % \vspace{-0.5cm}
    \label{fig:rq1_fas}
\end{figure}

\paragraph{Example 1.} Here we first take a SVHN \& LeNet5 pre-trained model as an example to illustrate the \fas introduced above. For the pre-trained model, there are 6 different feature maps in the first layer. Then we feed the labeled dataset $\mathcal{X}, \mathcal{Y}$ into the \fma with each feature map. \fas results are shown in~\cref{tab:example_fac}, we can observe that for different feature maps in the model's first layer, the \fas will be largely different. For instance, the 3th feature map is the most robust compared with other feature maps, which obtain low \fas for different attack strengths. In contrast, the 1st and 6th feature maps are vulnerable to even a 3-step \fma attack, which indicates that the 1st and 6th feature maps are vulnerable and that the DNN developers need to enhance them to avoid errors in safety-critical scenarios.
Then we further illustrate other combinations to evaluate the feature map's vulnerability.
As shown in~\cref{fig:rq1_fas}, the distribution of \fas is consistent across different strengths of the FMA, datasets, and models, indicating that the vulnerability of feature maps is an intrinsic characteristic of pre-trained models.
Additionally, the experiment results reveal that there are particularly vulnerable feature maps among those learned in a layer of the model~(e.g., the 12th feature map in~\cref{fig:subfig1-2}), attacking it can easily cause the model has incorrect behavior. 
These feature maps are what DNN developers need to pay extra attention to, as test cases that can induce these feature maps into error can more easily cause the model to misclassify. 
% The \fas metric is also used to test the model's deeper layer as in~\cref{fig:deepanalysis2}, revealing that there are also vulnerable feature maps as a model's depth increases, and the vulnerability of the feature map will increase as a model's depth increases.

\begin{table}[t]
    \centering
    \small
    \begin{tabular}{c|cccc}
    \toprule
         Feature Map&\multicolumn{4}{c}{Mutation Strength}  \\
         Index&I&II&III&IV\\
         \midrule
         1&4.88e-4&1.93e-3&4.29e-3&7.52e-3\\
         2&1.89e-4&7.34e-4&1.58e-3&2.70e-3\\
         3&4.97e-8&1.91e-7&4.16e-7&7.40e-7\\
         4&2.39e-4&9.31e-4&2.03e-3&3.52e-3\\
         5&1.83e-4&7.33e-4&1.64e-3&2.91e-3\\
         6&3.16e-4&1.26e-3&2.81e-3&4.96e-3\\
         \bottomrule
    \end{tabular}
    \caption{A example of \fvs in LeNet5 pre-trained model}
    \label{tab:example_fvs}
\end{table}

\begin{figure}[t]
    \begin{subfigure}[b]{0.45\linewidth}
        \subcaption[short for lof]{SVHN\&LeNet-5}
        \includegraphics[width=\linewidth]{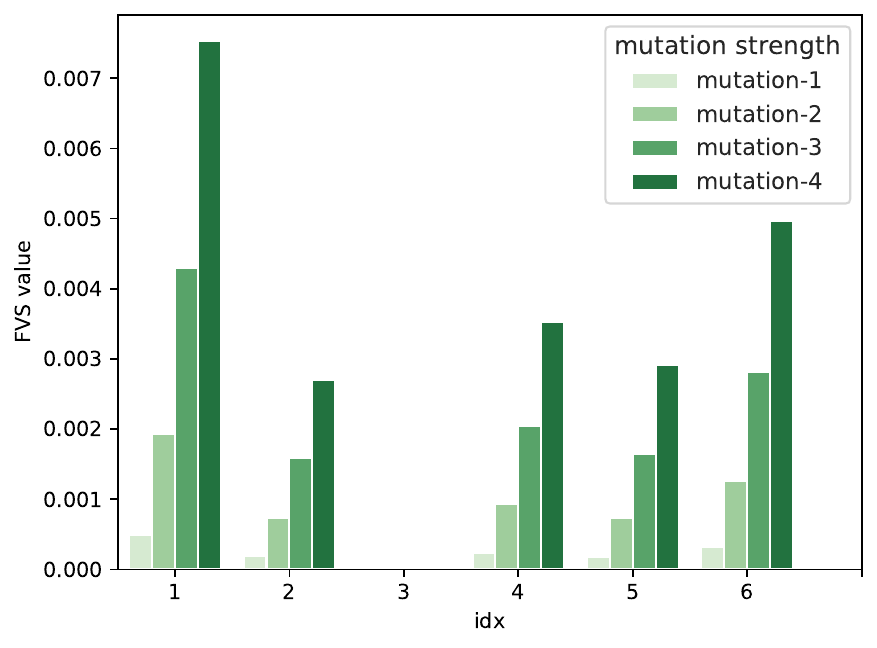}
        \label{fig:subfig2-1}
    \end{subfigure}%
        \begin{subfigure}[b]{0.45\linewidth}
        \subcaption[short for lof]{CIFAR10\&ResNet20}
        \includegraphics[width=\linewidth]{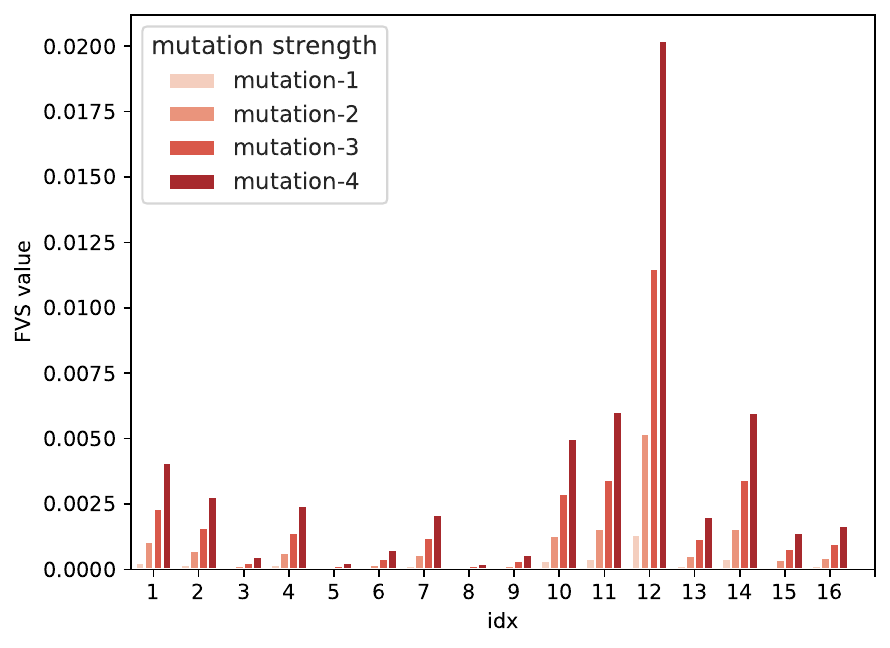}
        \label{fig:subfig2-2}
    \end{subfigure}%
  \caption{FVS value of different feature maps. The darker the color indicates the greater the degree of data perturbation, which is aligned with~\cref{fig:rq1_fas}. We use mutation to generate test cases here to be consistent with the selection experiment. Mutation refers to benign augmentation, whose strength is increased by increasing augmentation parameters' value range (as mentioned in~\cref{tab:setup}).
  }
  \label{fig:rq1_fvs}
\end{figure}

\begin{algorithm}
    \caption{Test Case Selection Pipeline}\label{algo:selection}
    \SetKwInput{Vars}{Variables}
    \SetKwProg{Fn}{Function}{:}{}
    \DontPrintSemicolon
    \SetNoFillComment
    \SetKwFunction{Fun}{Fun}
    \SetKwInOut{KOutput}{output}
    \KwIn{
     $x\in\mathcal{X}: \text{original test cases}\; x'\in\mathcal{X'}: \text{benign mutation test cases}\; f_{\theta}: \text{DNN to be tested}\; N: \text{The number of test cases to be selected}\; K: \text{TopK vulnerable feature map}$
     }
     \KOutput{$\mathcal{X_{s}}$: selected test cases}

     \Fn{FVS-Selection($f_{\theta}$, $\mathcal{X}$, $\mathcal{X}'$, $N$, $K$)}
     {
     \tcc{Detect TopK Vulnerable Feature Map}
     $FM$: Initialize vulnerable feature map list\\
    $FM \leftarrow f_{\theta}[\text{argsort}(\{FVS(f_{\theta}^c)\}_{\text{for each }} f_{\theta}^c \in f_{\theta})][-K:]$\\
    \tcc{Test Case Selection Begins}
    $\fvs\_array = []$\\
    \For{each $x_{i}, x_{i}' in \mathcal{X}, \mathcal{X}'$}{
    $FM\_score = 0$\\

    \For{each $f_{\theta}^{c} \in FM$}{
        $FM\_score += \fvs(f_{\theta}^{c}(x_{i}),f_{\theta}^{c}(x_{i}'))$
    }
    $\fvs\_array.append(FM\_score)$
    
    }
    $\mathcal{X_{s}} = \mathcal{X}[\text{argsort}(\fvs\_array)][-N:]$\\
    \Return $\mathcal{X_{s}}$
    }
\end{algorithm}

\subsection{\textbf{F}eature map \textbf{V}ulnerability \textbf{S}core}\label{sec:fvs}

As discussed in~\cref{sec:fac}, \fas will need the ground-truth label to calculate the score, which means once the developers only have unlabelled data, they can not use \fas to detect vulnerable feature maps.
To solve this problem, we introduce \fvs, another metric to detect vulnerable feature maps which measures the feature maps' vulnerability by directly computing the feature map output's difference between the test cases and its mutation examples~(i.e., feature map output's difference between $x$ and $x'$).
Formally, given a feature map $f_{\theta}^{c}$ in the model, a test set $\mathcal{X} = \{(x_i) \mid i = 1, 2, \dots, N\}$ and its corresponding mutated test set $\mathcal{X}' = \{(x_i') \mid i = 1, 2, \dots, N\}$ generated by a user-defined mutation method (e.g., rotation, shear, blur, and FMA), the \fvs is defined as:
% \begin{equation}
% \begin{split}
$$
\fvs(f_{\theta}^{c}) = \frac{1}{N}\sum_{i=0}^{N}(\Vert f_{\theta}^{c}(x) -  f_{\theta}^{c}(x')\Vert ^2)
$$
% &FVS( f_{\theta}^{c}(x), f_{\theta}^{c}(x') ) = \Vert 1 -  \frac{f_{\theta}^{c}(x')}{f_{\theta}^{c}(x)}\Vert ^2 \\
% & FVS( f_{\theta}^{c}(x), f_{\theta}^{c}(x') ) = \Vert f_{\theta}^{c}(x) -  f_{\theta}^{c}(x')\Vert ^2
% \nonumber
% \end{split}
% \end{equation}
% \vspace{-0.6cm}

where $f_{\theta}^{c}(x_{i})$ is the output of $x_{i}$ in the $c$-th feature map of the model $f_{\theta}$. 
The FVS for a feature map ranges between 0 and $\infty$, where a value of 0 means that the feature map will not be affected by the perturbation, which indicates that this feature map is not vulnerable. A value large than 0 means it will be affected by the perturbation, and the larger FVS for the feature map, the more vulnerable it is.

\paragraph{Example 2.}

In this part, we take the same SVHN \& LeNet5 pre-trained model as in \cref{sec:fac} to illustrate the effectiveness of \fvs introduced above. Specifically, for dataset $\mathcal{X}$ we use mutation methods in~\cref{tab:setup} to generate $\mathcal{X'}$, then we feed these cases into the model to calculate the \fvs for each feature maps in the first layer. Results are shown in~\cref{tab:example_fvs}, we can observe that the 3rd feature map has low \fvs value, while the 1st and 6th have high \fvs values, which means the 3rd feature map is robust while the 1st and 6th are vulnerable for the perturbation. Observations are consistent with the~\cref{tab:example_fac}, indicating the interchangeability between \fvs and \fas in label-agnostic scenarios. Then we further illustrate other combinations to evaluate the \fvs.
Results shown in~\cref{fig:rq1_fvs}, \fvs effectively captures the most vulnerable feature maps in the model. 
We find that FVS and FAS have the same distribution under the same model, which means we can use FVS to replace FAS in label-agnostic scenarios~(i.e., unlabeled dataset).

\begin{algorithm}\label{algo:fuzz}
    \caption{Fuzzer Pipeline}\label{algo:fuzz}
    \SetKwInOut{KOutput}{output}
    \SetKwProg{Fn}{Function}{:}{}
    \SetKwFunction{Fun}{Fun}
    \DontPrintSemicolon
    \SetNoFillComment
    \KwIn{$f_{\theta}$: pretrained DNN model,
    $x\in\mathcal{X}$: test cases;
    $\epsilon$: the fuzzing boundary,
    $\alpha$: and fuzzing step size,
    $steps$: the maximum number of steps to fuzz for a test case,
    $K$: TopK vulnerable feature map
    }
    \KOutput{$\mathcal{X_{s}}$: fuzzing cases generated by \fma-Fuzzer}
    \Fn{\fma-Fuzzer($f_{\theta}$,$\mathcal{X}$,$\epsilon$,$\alpha$,$steps$,$K$)}{
     \tcc{Detect TopK Vulnerable Feature Map}
     
     $FM$: Initialize vulnerable feature map list
     
    $FM \leftarrow f_{\theta}[argsort([FVS(f_{\theta}^c) \ for\  each \ f_{\theta}^c \in f_{\theta}])[-K:]]$
    
    \tcc{FMA-Fuzzer Begins}
    \For{each $x\in\mathcal{X}$}{
     \For {each $f_{\theta}^{c} \in FM$ }
     {
     $x' = FMA(f_{\theta}^c, x, steps, \alpha, \epsilon)$
     \If{$f_{\theta}(x^{'})!=y$}{
     $\mathcal{X_{s}}$.append($x^{'}$)
     }}
     }
     \Return $\mathcal{X_{s}}$
     }
\end{algorithm}

% \vspace{-0.5cm}
\subsection{Enhancing the DNN with \xxx}

\subsubsection{Test Case Selection From Unlabeled Data}\label{sec:selection}

Both testing and repairing the DNN-driven system rely on manually labeled data. While collecting a massive amount of unlabeled data is usually easy to achieve, the cost of manual labeling is much greater. For data with strong expertise knowledge~(e.g., medical data), it is unrealistic to blindly label all collected data. Therefore, DNN test case selection is crucial to select valuable data, reducing the labeling cost.

To select valuable test cases from unlabeled data, we propose a \fvs-guided test case selection method, which select test cases with high \fvs value. Algo~\ref{algo:selection} demonstrates the workflow of our \fvs-guided test case selection. 
Specifically, we first feed the unlabeled test cases and its benign mutation cases into the $FVS-Selection()$, then we will use $\mathcal{X}$ and $\mathcal{X}'$ to obtain the vulnerable feature map list~(Algo 2 line 3). After obtain the vulnerable feature map list, we will begin to select test cases. For each input $x$ and its mutation case $x'$, we will calculate the $FM\_score$ and then return the $\mathcal{X_{s}}$ that have larger $FM\_score$ than others.
% Specifically, we first compute the \fvs between each original test case $x$ and its mutation case $x'$ in vulnerable feature maps~(Algo~\ref{algo:selection} line 5-9). Then N test cases with the highest \fvs values will be returned~(Algo~\ref{algo:selection} line 10-11).

\subsubsection{Data augmentation by Fuzzing}\label{sec:fuzzing}

Another important application of DeepFeature is to generate data to increase the dataset capacity when the dataset is limited. Specifically, as one of the main elements for training DNN models, the dataset's size seriously affects the trained model's performance. We apply DeepFeature to the data augmentation task to solve this problem by using our FMA-Fuzzer to expand the dataset. \cref{algo:fuzz} presents the details of our proposed \fma-Fuzzer.
The inputs include the model $f_{\theta}$, the original test cases $\mathcal{X}$, the fuzzing boundary $\epsilon$, the maximum number of steps to fuzz for a test case $step$, and fuzzing step size $step\_size$. For each test case $x\in\mathcal{X}$, \fma-Fuzzer generates multiple samples corresponding to multiple vulnerable feature maps~(Alg~\ref{algo:fuzz}). Specifically, for each test case $x$, we iteratively generate perturbations targeting each feature map by maximizing the \fvs value~(Alg \ref{algo:fuzz} line 5-9). We first add a random perturbation to generate the initial $x^{'}$ as the first step. As we need to make sure the $loss$, formulated as:
$$
loss = \fvs(f_{\theta}^{c}(x'),f_{\theta}^{c}(x))
$$
starts at a non-zero value to obtain a non-zero gradient. The generated sample will be then added to the fuzzing list $\mathcal{X_{s}}$ if the model misclassifies it.

%% file: evaluation.tex
\section{Evaluation}

We evaluate \xxx and answer the following research questions.

\begin{itemize}
    % \item[\textbf{RQ1}] How effective is \xxx's metrics?
    % Can \xxx's metrics discover vulnerable feature maps in DNNs?
    \item \textbf{RQ1:} How effective is \xxx's test case selection?
    \item \textbf{RQ2:} How effective is \xxx's fuzzing algorithm?
    \item \textbf{RQ3:} Can \xxx behave stable under different settings?
\end{itemize}

\subsection{Experiment Setup}
% Our evaluation was done on a GPU server equipped with two twenty-core CPUs and four NVIDIA RTX2080Ti graphic cards. 

\paragraph{\textbf{Datasets and Models.}}
We adopt four widely used image classification benchmark datasets for the evaluation (i.e., MNIST~\cite{deng2012mnist}, Fashion MNIST~\cite{xiao2017/online}, SVHN~\cite{Netzer2011ReadingDI}, and CIFAR10~\cite{Krizhevsky09learningmultiple}), which are most commonly used datasets in deep learning testing~\cite{pei2017deepxplore,DeepMutation++,deepgini,ma2018deepgauge,Xie2018DeepHunterHD,Tian2018DeepTestAT,kim2019guiding,Gerasimou2020ImportanceDrivenDL,ismeanful,ADAPT,gao2022adaptive,Wang2021RobOTRT}. Table \ref{tab:dataset} presents the detail of the datasets and models. The MNIST~\cite{deng2012mnist} dataset is a large collection of handwritten digits. It contains a training set of 60,000 examples and a test set of 10,000 examples. The CIFAR-10~\cite{Krizhevsky09learningmultiple} dataset consists of 60,000 32x32 color images in 10 classes, with 6,000 images per class. There are 50,000 training images and 10,000 test images in CIFAR-10. Fashion~\cite{xiao2017/online} is a dataset of Zalando's article images—consisting of a training set of 60,000 examples and a test set of 10,000 examples. Each example is a 28x28 grayscale image associated with a label from 10 classes. SVHN~\cite{Netzer2011} is a real-world image dataset that can be seen as similar in flavor to MNIST (e.g., the images are of small cropped digits). SVHN is obtained from existing house numbers in Google Street View images. The models we evaluated include LeNet~\cite{lecun1998gradient}, VGG~\cite{simonyan2014very}, and ResNet~\cite{he2016deep}, which are also commonly used in deep learning testing tasks~~\cite{pei2017deepxplore,DeepMutation++,deepgini,ma2018deepgauge,Xie2018DeepHunterHD,Tian2018DeepTestAT,kim2019guiding,Gerasimou2020ImportanceDrivenDL,ismeanful,ADAPT,gao2022adaptive,Wang2021RobOTRT}. 

\paragraph{\textbf{Test Case Generation.}}
We follow the prior data simulation~\cite{Tian2018DeepTestAT,Ma2018DeepMutationMT, gao2022adaptive} to generate realistic test cases. Specifically, we use seven well-used benign mutations~(i.e., shift, rotation, scale, shear, contrast, brightness, and blur) to generate the test case with its original label. The configurable parameters of the mutation are shown in~\cref{tab:setup}. We do not choose adversarial attack~(e.g., FGSM, PGD, and BIM) to generate test cases because these data can not represent data collected from the real-world scenario and may lead to unreliable conclusions~\cite{li2019structural}.
During the test case generation, for each test data in the dataset, we randomly select one benign data augmentation from our seven augmentations to mutate a test case with its original label. When the original test size is 10,000, we will generate the same size of test cases.

\paragraph{\textbf{Fuzzing Case Generation.}}A branch of existing works~\cite{ADAPT,deepgini,tensorfuzz,pei2017deepxplore,Guo2018DLFuzzDF,DeepMutation++} have been proposed for generating fuzzing cases. To show the effectiveness of \xxx, we select the state-of-the-art neuron coverage-guided fuzzing method~(i.e., ADAPT~\cite{ADAPT}) and DeepMutation++~\cite{DeepMutation++} as our baseline. Compared with other existing works, ADAPT integrates with 29 neuron characteristics~(e.g., NAC), while others only have one to four different characteristics, which are even contained in ADAPT. The DeepMutation++ is also intergrate multiple coverage guided fuzzing metrics, e..g., Neuron-Level~(NAC), and layer-level~(e.g., TKNC). We believe evaluate \xxx's fuzzing module with ADAPT and DeepMutation++ can sohws \xxx's effectiveness. 
% We also notice~\citet{DeepHyperion_ISSTA_2021} propose DeepHyperion that use illumination search to generate test cases and obtain the SOTA performance. To compared with it, we also select it as our baseline to evaluate \xxx's fuzzer effectiveness.
We also notice that DeepHyperion~\cite{DeepHyperion_ISSTA_2021} generates test cases from the feature map level, so we chose it as one of our baseline.

\paragraph{\textbf{Test Case Selection}}
Multiple test case selections have been proposed by recent works~(e.g., NC-guided~\cite{pei2017deepxplore,Tian2018DeepTestAT,ma2018deepgauge,Gerasimou2020ImportanceDrivenDL,Xie2022NPCNP}, Priority~\cite{deepgini,Wang2021PrioritizingTI}, Active learning guided~\cite{gao2022adaptive,Ren2020ASO}, robust-guided~\cite{Wang2021RobOTRT,madry2017towards, Chen2020PracticalAE}, SA-guided~\cite{kim2019guiding}).
However, robust-guided selection~\cite{Wang2021RobOTRT,madry2017towards,Chen2020PracticalAE} focus on model robustness, as mentioned by~\citet{Zhang2019TheoreticallyPT}, there is a trade-off between the accuracy and robustness, which means using these~(e.g., RobOT~\cite{Wang2021RobOTRT}, and PACE~\cite{Chen2020PracticalAE}) strategies to increase model robustness, the accuracy will decrease. So we will not choose these strategies as our baselines.

To show the effectiveness of \xxx, we use the most famous metric~(i.e., NAC) and its fine-grained version~(i.e., KMNC) proposed by DeepGauge~\cite{ma2018deepgauge}. To evaluate \xxx with SOTA coverage metrics, we take NPC~\cite{Xie2022NPCNP} as our baseline. Compared with DC/MC, it can evaluate the DNN decision logic flows for each connected layer directly, which can reduce the overhead of DC/MC. Since \xxx's test case selection is a prioritization technique~(i.e., priority the test cases with certain rules and return cases with larger priority), we take DeepGini, the SOTA open-sourced prioritization technique as our baseline. Compared with PRIMA~\cite{Wang2021PrioritizingTI}, it only need $1/100$ times to select test cases. To compare \xxx with SA metrics, we use DSA, which is proposed by~\citet{kim2019guiding}, to evaluate \xxx's effectiveness. Then to evaluate \xxx's effectiveness with the current SOTA active learning guided selection strategies, we use ATS~\cite{gao2022adaptive} as our baseline. Finally, we also use Random Selection~(RS) as a natural baseline, which can help us evaluate whether a selection method is effective. 
The configurable parameters of the selection strategies are shown in~\cref{tab:config}.
% To compare with coverage-guided test selection strategies, we select four popular neuron coverage criteria~(i.e., NAC~\cite{pei2017deepxplore}, KMNC, TKNC, and SNAC~\cite{ma2018deepgauge}) as our baseline to guide the test case selection. 
% Specifically, for any coverage criteria in our baseline, it follows an additional greedy algorithm to select the next test case according to the feedback from the previously selected test case (i.e., select the test case that covers the maximum number of uncovered areas of the given coverage criteria).
% \paragraph{\textbf{Prioritization test case Selection}}
% we conduct our experiments with DeepGini~\cite{deepgini}, which is a widely-used prioritization selection method.
% It takes the use of the Gini coefficient to measure the likelihood of test case $i$ being misclassified.

\begin{table}[t]
    \centering
\small
    \begin{tabular}{c|c c}
    \toprule
         Transformations&Parameters& Parameter ranges  \\
    \midrule
         Shift&$(s_x, s_y)$ &[0.05, 0.15]\\
         Rotation&$q~(degree)$&[5, 25]\\
         Scale&$r~(ratio)$&[0.8,1.2]\\
         Shear&$s~(angle)$&[15, 30]\\
         contrast&$\alpha~(gain)$&[0.5,1.5]\\
         Brightness& $\beta~(bias)$&[0.5,1.5]\\
         Blur&$ks~(kernel size)$&\{3,5,7\}\\
    \bottomrule
    \end{tabular}
    \caption{Transformations and parameters used by \xxx for generating test cases.}
    % \vspace{-0.3cm}
    \label{tab:setup}
\end{table}

\begin{table}[t]
\small
    \centering
    \begin{tabular}{c|c c}
    \toprule
         Criteria&Parameters &Parameter Config \\
         \midrule
         Random&- &-\\
         NAC&t~(threshold)&0.5\\
         KMNC&k~(k-bins)&1000\\
         NPC&$\alpha$&0.7\\
         DSA&$n$&1000\\
         Gini&None&None\\
         ATS&None&None\\
        \xxx&\multicolumn{2}{c}{K=5, step\_size=7, $\alpha=\epsilon/4=1/255$}\\
    \bottomrule
    \end{tabular}
    \caption{The parameter configuration of test case selection. The $step\_size =7$ is from~\cref{fig:rq1_fas}.}
    \label{tab:config}
    % \vspace{-0.7cm}
\end{table}

\begin{table*}[t]
    \centering
    % \small
    % \setlength{\tabcolsep}{2.5pt}
        \begin{tabular}{l|c|ccccccc|c|ccccccc}
        \toprule
        \multirow{2}{*}{Dataset(DNN)}&\multicolumn{8}{c|}{Select 5\% Test Cases}&\multicolumn{8}{c}{Select 10\% Test Cases}\\
        &Our&NAC&KMNC&NPC&DSA&Gini&ATS&RS&Our&NAC&KMNC&NPC&DSA&Gini&ATS&RS\\
        \midrule
        MNIST~(L-1)&\textbf{84.4}&22.6 &42.2 &20.2 &22.2 &61.8&58.7&21.3&      \textbf{78.0}&20.8 &38.7&20.2 &22.5&55.7&54.3&21.3\\
        MNIST~(L-5)&\textbf{66.2}& 18.4& 25.8& 21.6& 22.6&58.2&60.2&18.8&      \textbf{61.4}& 18.1& 20.8& 20.0& 21.3&50.5&52.3&18.8\\
        Fashion~(L-1)&\textbf{67.0}& 32.7& 35.9& 31.0& 31.1&57.8&53.3&31.2&    \textbf{63.8}& 30.9& 35.9& 31.0& 31.0&48.2&48.7&31.0\\
        Fashion~(R-20)&\textbf{79.8}& 21.1& 23.7& 30.5& 24.1&55.0&40.3&26.3&    \textbf{70.6}& 20.5& 22.0& 28.4& 27.1&48.7&35.2&26.2\\
        SVHN~(L-5)&\textbf{73.0}& 29.1& 28.8& 31.0& 31.1&53.2&47.8&28.8&       \textbf{67.6 }& 29.1 & 27.2 & 31.0 & 31.1&47.3&42.1&29.1\\
        SVHN~(V-16)&\textbf{63.6}&21.5 &16.2 &18.9 &24.5 &53.0&55.3&16.0&        \textbf{59.6}& 19.6&16.2 & 17.1&23.4&43.1 &48.7&16.0 \\
        CIFAR-10~(V-16)&\textbf{67.4} & 28.4 &  16.4&30.6  &43.6 &60.2&62.1&30.9&\textbf{60.2}& 29.4& 17.4&30.3 & 43.3&56.2&56.3&30.8\\
        CIFAR-10~(R-20)&\textbf{59.6}& 28.2& 24.6& 26.0& 27.4&50.4&50.2&28.8& \textbf{53.7}& 25.8& 21.8& 28.7& 27.4&45.0&47.1&28.9\\
        \midrule
        \multirow{2}{*}{Dataset(DNN)}&\multicolumn{8}{c|}{Select 15\% Test Cases}&\multicolumn{8}{c}{Select 20\% Test Cases}\\
        &Our&NAC&KMNC&NPC&DSA&Gini&ATS&RS&Our&NAC&KMNC&NPC&DSA&Gini&ATS&RS\\
        \midrule
        MNIST~(L-1)&\textbf{70.7}& 21.6& 35.3& 20.6& 23.6&50.1&47.6&21.3& \textbf{62.1} & 22.0 & 35.7 & 20.9 & 23.5&45.4&41.5&21.3\\
        MNIST~(L-5)&\textbf{56.2}& 18.5& 19.3& 19.3& 21.0&46.2&40.2&18.7&\textbf{51.5} & 18.4 & 19.4 & 19.1 & 20.8&42.3&37.1&18.7\\
        Fashion~(L-1)&\textbf{58.4} & 30.5 & 34.2 & 31.0 & 31.0&44.4&45.1&31.21& \textbf{56.8}& 30.3& 34.6& 31.1& 31.0&39.4&40.1&31.3\\
        Fashion~(R-20)&\textbf{66.1}& 21.1& 22.2& 27.3& 27.8&47.2&30.3&26.2& \textbf{60.7} & 21.4 & 22.8 & 26.7 & 27.6&42.7&28.5&26.2\\
        SVHN~(L-5)&\textbf{64.7}& 29.1& 27.5& 30.9& 31.0&41.3&38.9&28.9& \textbf{62.3} & 29.1 & 28.3 & 30.9 & 31.0&35.4&33.1&28.9\\
        SVHN~(V-16)&\textbf{59.3} &18.6  &15.5  &17.0  &23.5 &34.7&43.2&15.9& \textbf{58.6}&18.2&15.6 &16.6 & 23.1&28.9&40.1&15.9\\
        CIFAR-10~(V-16)&\textbf{54.4} & 27.0 & 17.7 & 30.6 & 42.8&51.9&50.5&30.8&\textbf{50.8} & 29.8 & 18.3 &30.7 &42.3 &48.5&46.1&30.8 \\
        CIFAR-10~(R-20)&\textbf{48.3} & 25.6 & 21.5 & 28.8 & 27.2&40.6&42.7&28.9& \textbf{47.9}& 26.4& 20.7& 28.5& 26.4&35.4&36.9&28.9\\
        \bottomrule
        \end{tabular}

    \caption{Fault Detection Rate of \xxx and baselines.}
    \label{tab:num_faults}
    % \vspace{-0.3cm}
\end{table*}

\begin{table*}[]
    \centering
    % \small
    % \setlength{\tabcolsep}{2pt}
    \begin{tabular}{l|c|ccccccc|c|ccccccc}
        \toprule
        \multirow{2}{*}{Dataset(DNN)}&\multicolumn{8}{c|}{Select 5\% Test Cases}&\multicolumn{8}{c}{Select 10\% Test Cases}\\
        &Our&NAC&KMNC&NPC&DSA&Gini&ATS&RS&Our&NAC&KMNC&NPC&DSA&Gini&ATS&RS\\
        \midrule
        MNIST~(L-1)&\textbf{8.01}&7.07 &7.31 &7.10 &7.10& 7.54&7.41&7.09 &\textbf{8.18}&7.14 &7.42 &7.14 &7.15&7.91& 7.82&7.14\\
        MNIST~(L-5)&\textbf{8.65}&7.04 &7.24 &7.03 &7.04& 7.57&7.21&7.04 &\textbf{9.27}&7.06 &7.31 &7.08 &7.07&8.21 &7.24&7.06\\
        Fashion~(L-1)&\textbf{6.03}&4.95 &5.01 &4.98 &4.97&5.31& 5.73& 4.97&\textbf{6.44}&5.04 &5.11 &5.05 &5.03&5.77& 6.23&5.04\\
        Fashion~(R-20)&\textbf{6.17}&6.08 &6.10 &6.09 &6.10&6.13&6.15 &6.08 &\textbf{7.39}&6.14 &6.23 &6.12 &6.16&6.37& 6.42&6.14\\
        SVHN~(L-5)&\textbf{7.10}&3.86 &3.94 &3.85 &3.86&4.21&4.23 &3.85 &\textbf{7.49}&4.12 &4.22 &4.13 &4.14&5.21&5.17 &4.13\\
        SVHN~(V-16)&\textbf{3.77}&2.11 &2.41 &2.12 &2.12&2.57&2.81 &2.11 &\textbf{4.08}&2.40 &2.51 &2.38 &2.39&2.92&3.03 &2.38\\
        CIFAR-10~(V-16)&\textbf{7.17}&1.51 &1.61 &1.50 &1.51&2.12&2.05 &1.50 &\textbf{8.21}&1.82 &1.91 &1.81 &1.82&2.73&2.33 &1.81\\
        CIFAR-10~(R-20)&\textbf{4.43}&2.35 &2.47 &2.35 &2.34&2.79&2.81 &2.35 &\textbf{5.13}&2.44 &2.54 &2.43 &2.41&3.02&3.15 &2.43\\
        \midrule
        \multirow{2}{*}{Dataset(DNN)}&\multicolumn{8}{c|}{Select 15\% Test Cases}&\multicolumn{8}{c}{Select 20\% Test Cases}\\
        &Our&NAC&KMNC&NPC&DSA&Gini&ATS&RS&Our&NAC&KMNC&NPC&DSA&Gini&ATS&RS\\
        \midrule
        MNIST~(L-1)&\textbf{8.21}&7.20 &7.49 &7.21 &7.20 &8.03&7.93&7.20 &\textbf{8.24}&7.26 &7.53 &7.26 &7.27&8.07&8.01 &7.26\\
        MNIST~(L-5)&\textbf{9.42}&7.07 &7.37 &7.09 &7.08&8.43 &7.25&7.08 &\textbf{9.49}&7.09 &7.41 &7.10 &7.08&8.62 &7.25&7.08\\
        Fashion~(L-1)&\textbf{6.75}&5.10 &5.21 &5.09 &5.08&6.01 &6.31&5.09 &\textbf{7.01}&5.12 &5.24 &5.12 &5.13&6.15&6.47 &5.12\\
        Fashion~(R-20)&\textbf{7.70}&6.26 &6.31 &6.27 &6.27 &6.52&6.60&6.26 &\textbf{7.79}&6.31 &6.37 &6.32 &6.31&6.63&6.63 &6.32\\
        SVHN~(L-5)&\textbf{7.80}&4.38 &4.45 &4.36 &4.37 &5.77&5.48& 4.37 &\textbf{7.98}&4.61 &4.72 &4.61 &4.59&5.92&5.69 &4.61\\
        SVHN~(V-16)&\textbf{4.21}&2.57 &2.65 &2.58 &2.57&3.16 &3.17&2.57 &\textbf{4.25}&2.80 &2.92 &2.82 &2.79&3.32 &3.40&2.81\\
        CIFAR-10~(V-16)&\textbf{8.30}&2.02 &2.22 &2.04 &2.03&3.21&2.51 &2.04 &\textbf{8.33} &2.49 &2.57 &2.49 &2.49&3.37 &2.77&2.49\\
        CIFAR-10~(R-20)&\textbf{5.77}&2.55 &2.71 &2.55 &2.56&3.47 &3.41&2.56 &\textbf{5.81}&2.71 &2.77 &2.72 &2.72&3.91 &3.76&2.71\\
        \bottomrule
    \end{tabular}
    \caption{Increase in DNN's accuracy~(\%) after repairing the DNN with test cases selected by DLS testing.}
    % \vspace{-0.7cm}
    \label{tab:repairing}
\end{table*}

\subsection{\textbf{RQ1:How effective is \xxx's test case selection?}}\label{sec:eval:selection}
\subsubsection{Fault Detection Effectiveness}
Similar to traditional software testing~\cite{Gligori2015PracticalRT,Zhang2018HybridRT,Legunsen2016AnES}, test case selection tries to find valuable test cases from a large pool of candidate unlabeled test set, which can reduce the cost of manual labeling time once the labeling resource is limited. For a given selection method, a selected test set that can trigger more faults means it could reveal more defects in the software. We take \textbf{F}ault \textbf{D}etection \textbf{R}ate~(FDR) as our evaluation metric to measure the effectiveness of \xxx's test case selection method. Specifically, the FDR is defined as follows:
$$FDR(X) = \frac{|X_{wrong}|}{|X|}$$
where $|X|$ denotes the size of the selected test cases, and $|X_{wrong}|$ represents the number of test cases being misclassified by DNN.

As shown in~\cref{tab:num_faults}, we compare the FDR of \xxx and our baselines at different test case selection rates~(5\%, 10\%, 15\%, and 20\%).
First, we found that NC-guided and test case selection methods~(i.e., NAC, KMNC, and NPC) have poor performance in fault detecting, and sometimes their FDR are very similar to RS's, which indicates that neuron coverage is not a proper metric to guided test case selection, and this is consistent with previous research~\cite{li2019structural,ismeanful}. 
And we also notice that DSA is also similar with RS's performance, which may because the surprise adaquacy is also not correlated with FDR or not correlated with the prediction confidence. Compared to the coverage-guided test selection baseline, \xxx, DeepGini, and ATS show a higher defect detection capability, while \xxx is still the best result for most combinations. Specifically, compared to coverage-guided and prioritization test case selection methods, \xxx can improve FDR by up to 49.32\% and 24.8\% in MNIST\&LeNet1, respectively. We believe that the main reason for \xxx having a higher FDR is that, as a feature-map-guided testing technique, \xxx can detect more types of defects that do not induce neuron-type defects, but feature-map-type defects (see~\cref{sec:eval:faulttype}).

\subsubsection{Repairing Effectiveness}
Finding faults and using them to enhance the DNN model by retraining is the ultimate purpose of DNN testing. To evaluate the repairing effectiveness of \xxx, we add selected test cases, with selection ratios ranging from 5\% to 20\%, to retrain the models, as listed in Table~\ref{tab:repairing}. For each dataset \& model combination, we use the same training hyperparameters~(e.g., epoch, optimizer settings) to perform a fair comparison. Specifically, all models are trained for 40 epochs, with the learning rate initialized to 0.001 and stepped down to one-tenth of the original in the 20th and 30th epochs. We use SGD as the optimizer with a Nesterov momentum of 0.99. Detailed accuracy improvement results are listed in \cref{tab:repairing}.
We achieved the highest model accuracy improvement across all datasets, models, and test case selection ratios~(e.g., In CIFAR 10 \& ResNet20 combination, \xxx can increase 5.81\% accuracy while baselines can only maximum increase 3.91\% accuracy), which indicates that \xxx can repair more faults in DNN models compared with our baselines. 

% \mybox{Finding 4: While not perfect, results from selection and repairing experiments shows that \xxx has a stronger defect detection capability than our baselines. \xxx can select more valuable test cases, with which we repair the model for further improvement in the model accuracy.}

\begin{figure*}[]
	% \begin{subfigure}{0.24\linewidth}
	% 	\centering
	% 	\includegraphics[width=1\linewidth]{diversity_mnist_lenet1.pdf}
	% 	\caption{MNIST \& LeNet1}
	% \end{subfigure}
	\centering
	\begin{subfigure}{0.23\linewidth}
		\centering
		\includegraphics[width=1\linewidth]{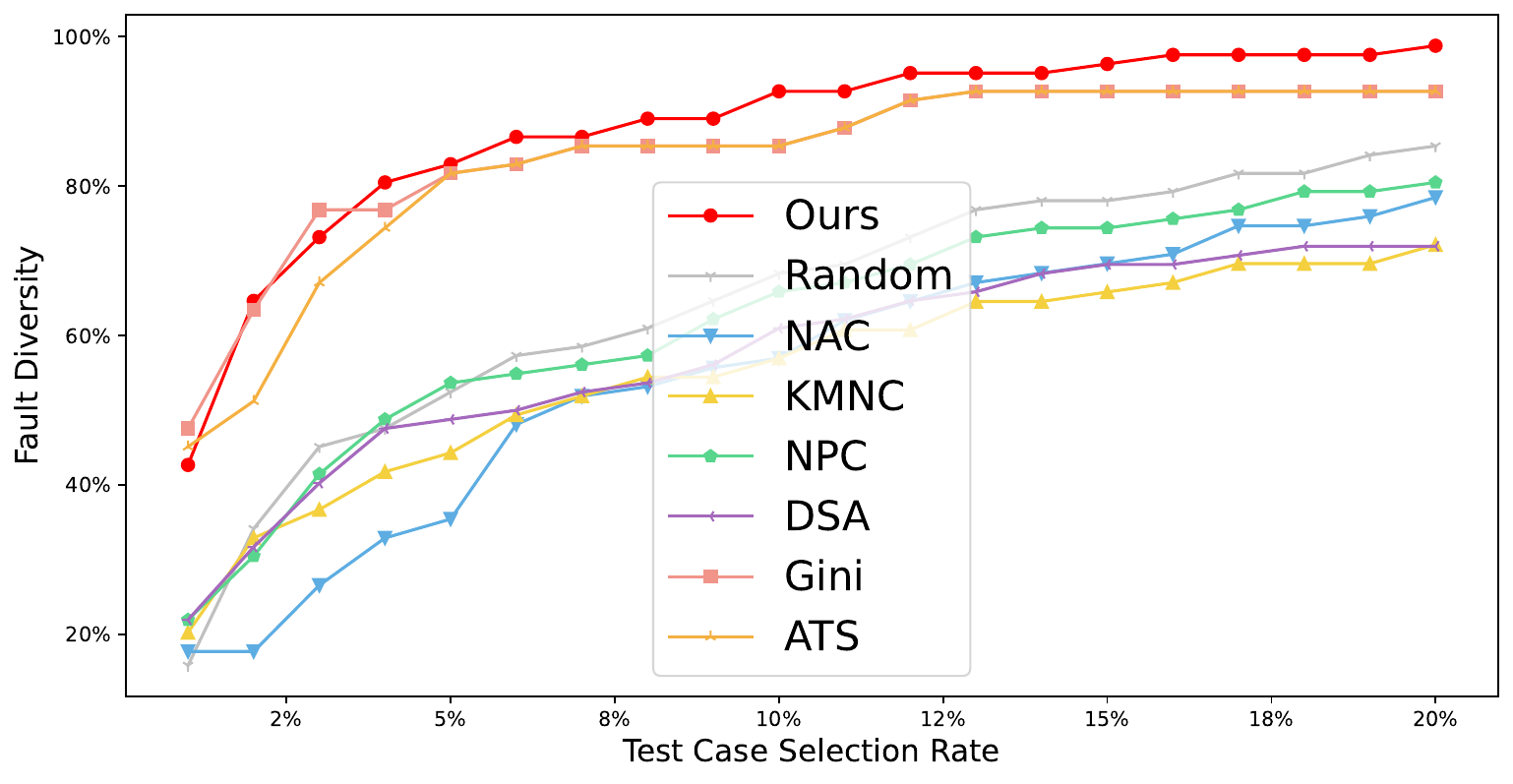}
		\caption{MNIST \& LeNet5}
	\end{subfigure}
	\begin{subfigure}{0.23\linewidth}
		\centering
		\includegraphics[width=1\linewidth]{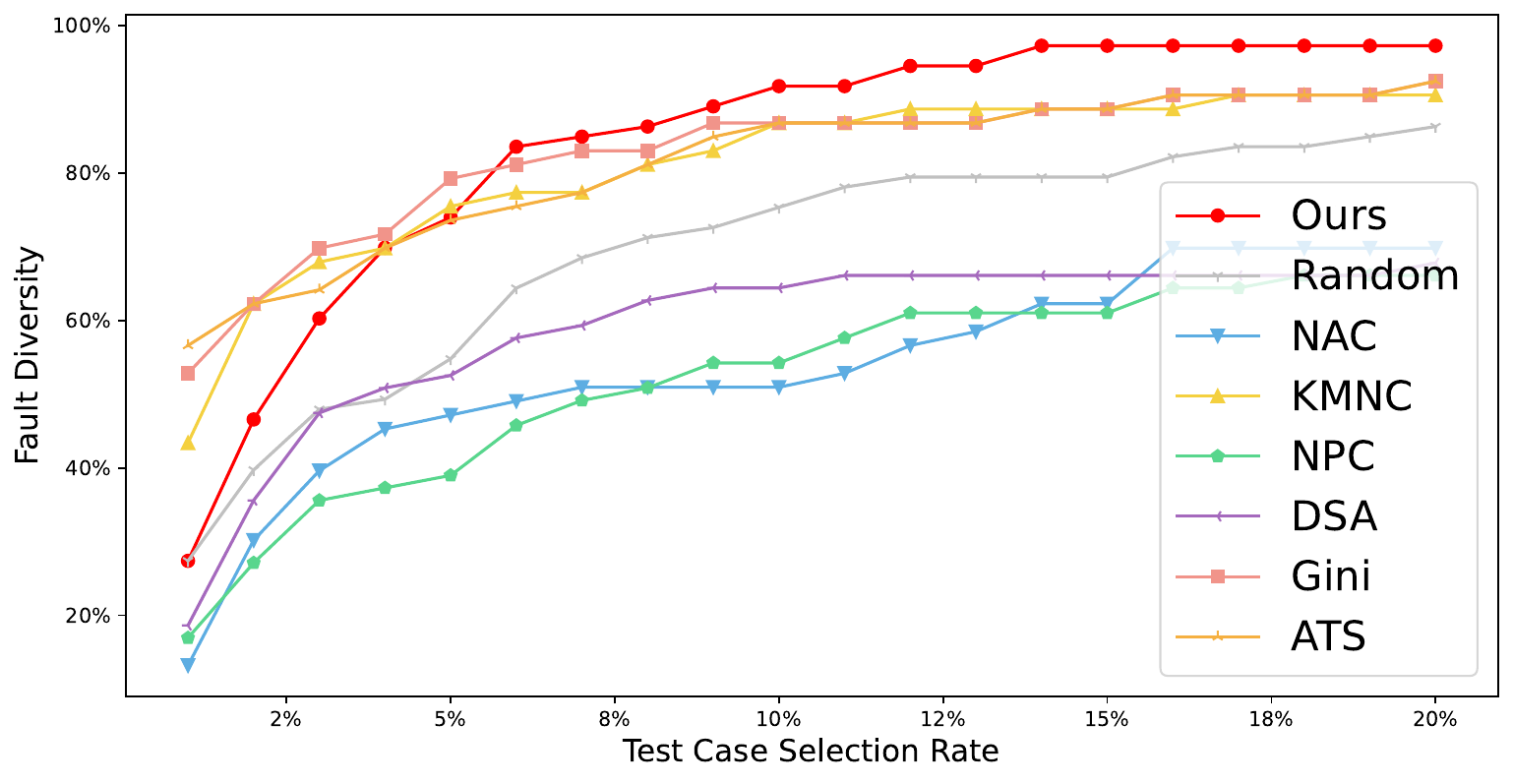}
		\caption{Fashion \& LeNet1}
	\end{subfigure}
		\begin{subfigure}{0.23\linewidth}
		\includegraphics[width=1\linewidth]{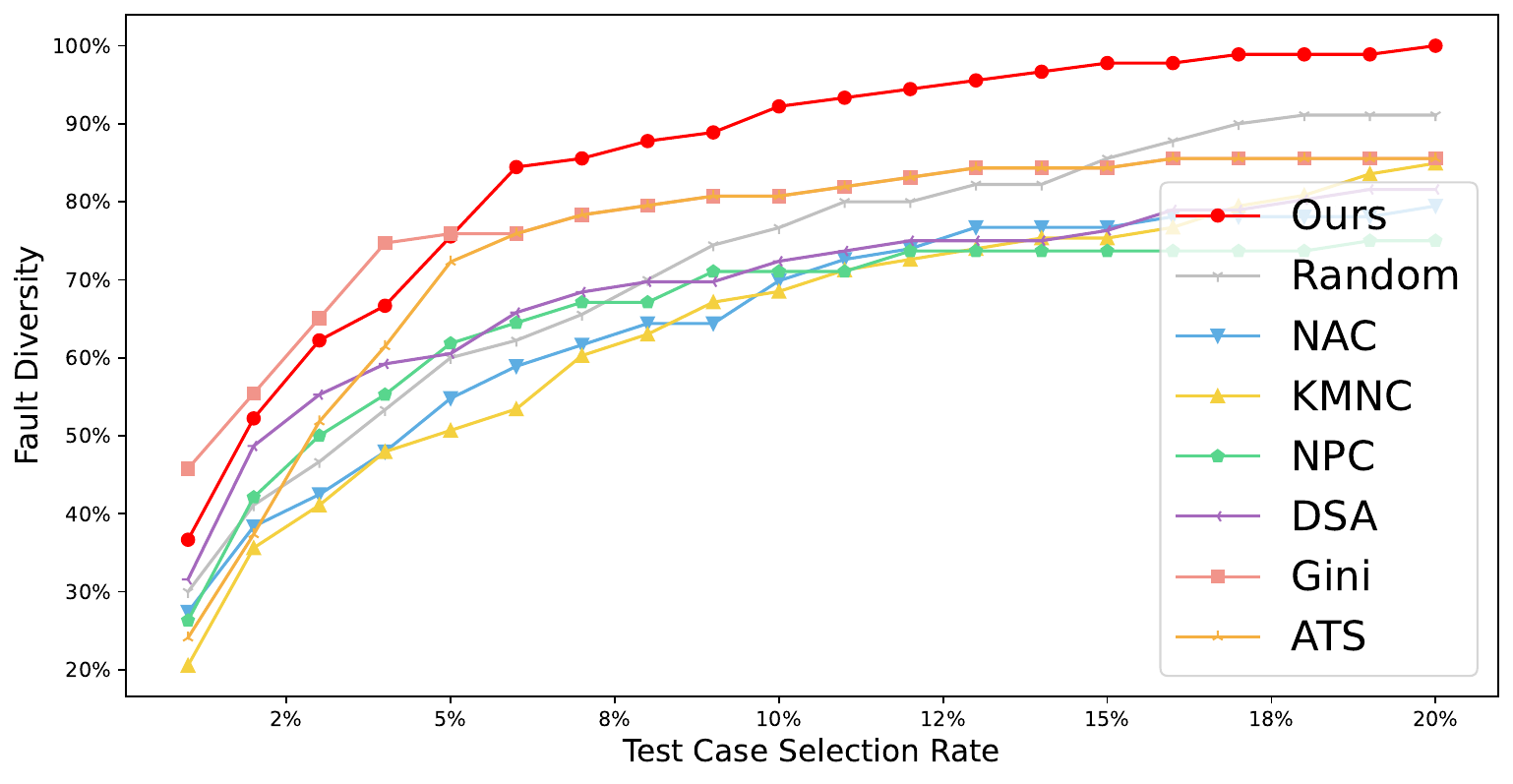}
		\caption{CIFAR10 \& ResNet20}
	\end{subfigure}
 	\begin{subfigure}{0.23\linewidth}
		\centering
		\includegraphics[width=1\linewidth]{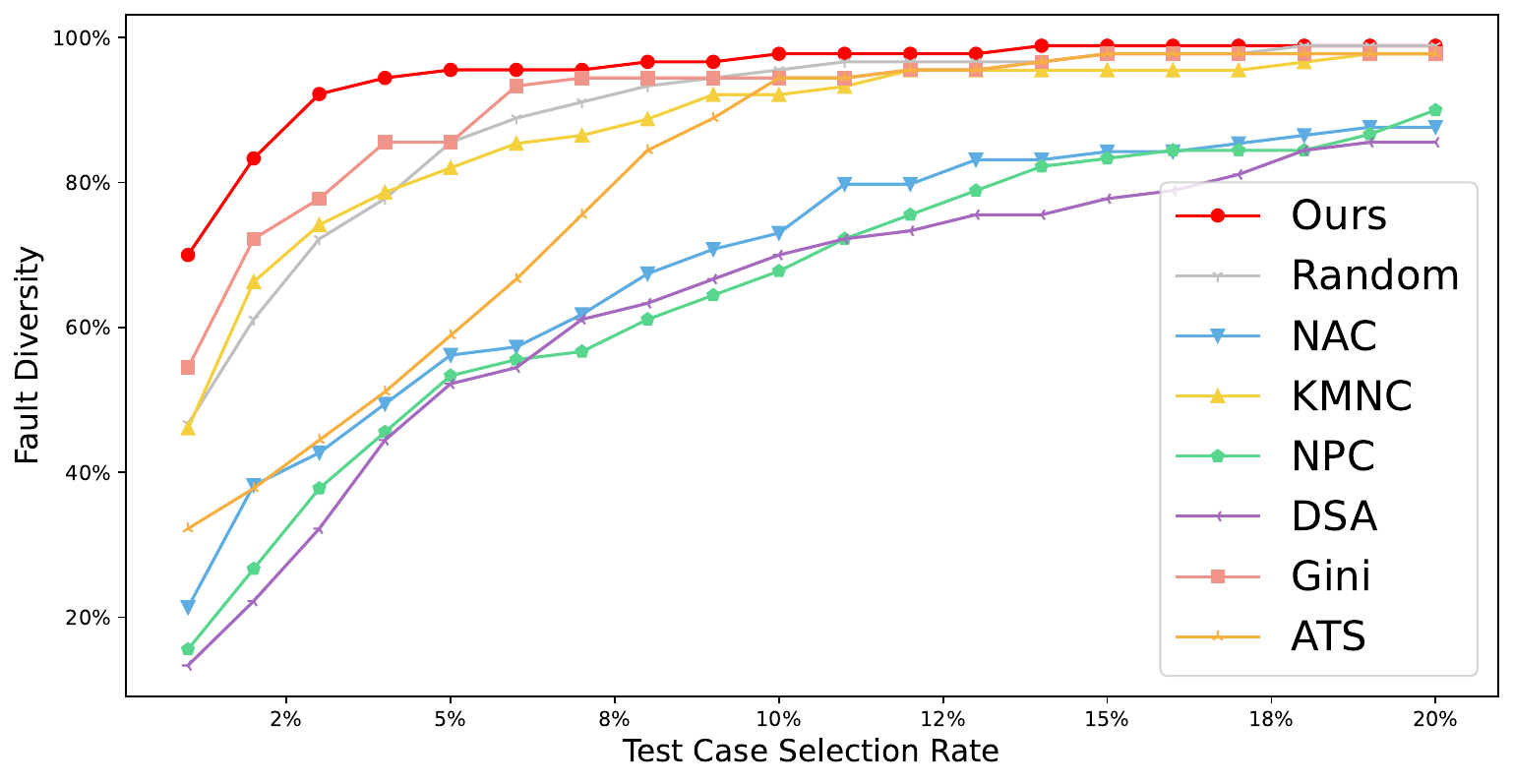}
		\caption{SVHN \& LeNet5}
	\end{subfigure}
	% \begin{subfigure}{0.24\linewidth}
	% 	\centering
	% 	\includegraphics[width=1\linewidth]{diversity_fashion_resnet20.pdf}
	% 	\caption{Fashion \& ResNet20}
	% \end{subfigure}
 
	% \centering
	% \begin{subfigure}{0.24\linewidth}
	% 	\centering
	% 	\includegraphics[width=1\linewidth]{diversity_cifar10_vgg16.pdf}
	% 	\caption{CIFAR10 \& VGG16}
	% \end{subfigure}
	% \begin{subfigure}{0.24\linewidth}
	% 	\centering
	% 	\includegraphics[width=1\linewidth]{diversity_svhn_vgg16.pdf}
	% 	\caption{SVHN \& VGG16}
	% \end{subfigure}
 % \vspace{-0.3cm}
	\caption{The cumulative sum of the fault types coverage rate~(FTCR) found by our selection method. We only report four combinations due to page limitaion. Please refer to~\cref{tab:auc} for quantitative results.}
 % \vspace{-0.5cm}
	\label{fig:diverse}
\end{figure*}

\subsubsection{Fault Type Detection Effectiveness}\label{sec:eval:faulttype}

In this part, we delve deeper into the reasons why \xxx has a higher FDR and can increase more in accuracy than the baselines.
Insights from traditional software testing found error-inducing inputs are very dense, inspiring DNN testing that detecting a greater variety of errors may also be as important as detecting more errors. 
We use the concept of fault type to answer this question. For a given test case $x$ that has been misclassified, its fault type is defined as:
$$
Fault\_Type(x) = (Predict(x) \rightarrow Label(x))
$$
where $Label(x)$ denotes the ground-truth label, and $Predict(x)$ denotes the DNN prediction. For instance, if the $Predict(x)$ is ``7'', while the $Label(x)$ is ``9'', then the fault type of $x$ is denoted as: $Fault\_Type(x) = 7 \rightarrow 9$.
For a typical classification dataset with 10 different categories, the number of possible fault types is $10\times9=90$. As the candidate test cases to be selected may not introduce all types of errors, we use the \textit{Fault Type Coverage Rate~(FTCR)}, the proportion of error types introduced by our selected test cases among the error types introduced by all samples, to quantify the ability to select diverse faults.

Plots of FTCR with the percentage of selected cases increasing from 1\% to 20\% are shown in~\cref{fig:diverse}. Our selection method achieved better fault diversity compared to random selection and baseline methods under the vast majority of dataset\&model combinations, indicating that DeepFeature can detect more types of faults in DNN,  which explains why \xxx has higher FDR and can increase more accuracy compared with baselines.
Second, we can see that in most of our experiments, the neuron coverage-driven selection stops increasing after the FTCR reaches 80\%. Even if we increase the selection rate after that, their FTCR still converges to 80\%. However, DeepFeature's FTCR continues to increase to 100\%, implying that some cases are not detected by neuron coverage-guided selection. In contrast, DeepFeature can detect them, implying that these cases will induce feature map types of errors.
% We also find that KMNC obtained the best results in the baseline approach in experiments with LeNet1 as the tested model, while TKNC obtained the highest diversity in experiments with VGG16, which happened to be the two models with the minimum and a maximum number of neurons. This phenomenon may indicate the preference and limitations of the two metrics for different model sizes. The other two metrics in the baseline~(NAC and SNAC) bring even worse diversity than random selection in all dataset \& model combinations. 

As shown in the \cref{tab:auc}, we also calculated the ratio of area under the curve~(RAUC) of FTCR to quantitatively demonstrate how well each approach can identify various defects. It can be seen that existing neuron-coverage metrics are generally weaker than even random selection in terms of detecting diverse defects. Homogeneous faults from neuron coverage guided selection make only a trivial contribution to follow-up repairing. Nevertheless, \xxx tries to select more types of faults, yielding greater accuracy improvement after retraining~(repairing) the model.

\mybox{Answer to RQ1: \xxx can select more diverse and valuable test cases, with which we repair the model for further improvement in the model accuracy.}

% \vspace{-0.6cm}

\subsection{\textbf{RQ2: How effective and efficient is \xxx's fuzzing algorithm?}}\label{sec:eval:fuzzer}

To answer this question, we compare \xxx with the SOTA fuzzing algorithm ADAPT and DeepMutation++~(DeepM++), and DeepHyperion~(DeepH). We run \xxx and baselines for the same period of time~(i.e., 5 minutes) to generate test cases. For the baselines' parameter, we follow the default settings in its original paper~(i.e., the fuzzing time for each case is limited to 1 second, and the coverage follows~\cref{tab:config}). Then we retrain the models with generated cases and re-evaluate them to compare the accuracy improvement brought by \xxx and baselines.

Table~\ref{tab:fuzz} shows our experiment results. We can see that for the same amount of time, \xxx can mutate more valuable fuzzing cases, which will be misclassified by the model, than baselines. Then, we also evaluate the quality of the fuzzing cases generated by baselines and ours. Considering baselines' fuzzing cases are less than \xxx, we randomly select the same number of cases from \xxx to make a fair comparison for the fuzzing effectiveness~(e.g., we only select 771 cases from \xxx in MNIST \& LeNet1 combination). We then retrain the model using these fuzzing cases. Experiment results reveal that \xxx brings greater improvement on model accuracy in the majority of dataset \& model combinations~(e.g., we improve the accuracy of MNIST\&LeNet1 into 93.45\%, while baselines only improve to 92.84\% and 92.88\%.), denoting that test cases fuzzed by \xxx are more valuable than baselines.

Then, we can also observe that compared with DeepH, \xxx can generate more misclassified cases 5 to 20 times, which is because DeepHyperion will use illumination search to generate test cases for each feature map in the model, which causes it becomes time-consuming. Then we can also observe that using the same number of cases to repair the model, \xxx can increase more accuracy compared with DeepH, which is because cases generated by \xxx are focused on repairing vulnerable feature maps, while cases generated by DeepH may fine-tune the feature maps that are robustness, which decrease the repairing efficiency.
% and then the repairing efficiency will decrease.

\mybox{Answer to RQ2: \xxx is more effective and efficient than the baseline approaches. }

\subsection{Can \xxx behave stable under different settings?}\label{sec:eval:sample}

The FDR experiment results in \cref{tab:num_faults} are based on the five most vulnerable feature maps across all datasets and models. However, a fixed number of feature maps may not generalize well to models of different sizes, so we also analyze how the number of vulnerable feature maps would affect the estimation of the \textit{value} of a test case and further affect the FDR. Specifically, for each experiment combination, we use the top-1, 5, 10, 15, 20, and 25 vulnerable feature maps to guide test case selection. 
Results are listed in \cref{tab:size}. 
% We find that the best choice of the number of vulnerable feature maps varies for different datasets \& model combinations. For instance, in CIFAR10 \& ResNet20 combination, when Top-1 vulnerable feature map is used to guide test case selection, \xxx can obtain the best FDR result compared with other Top-K~(e.g., 5, 10, 15, 20, and 25), which is because, in our CIFAR10 \& ResNet20 pre-trained model, there is a vulnerable feature map, whose \fvs~(\cref{fig:subfig1-2}) and \fas~(\cref{fig:subfig2-2}) are largely different with other feature maps. Another intriguing finding from our studies is that once the K is more than 5, \xxx may perform competitively, meaning that even if the DNN developers are unfamiliar with \xxx, they can still utilize our default setting~(i.e., Top-5 vulnerable feature maps) to test DNNs.
We find that once the number of selected vulnerable feature maps are larger than 5, the effectiveness of \xxx will become stable. For instance, when number from 5 to 25, \xxx's FDR only change from 62.10\% to 62.01\% in MNIST\&LeNet1 combination, and the RAUC only change from 87.97\% to 88.55\%, which indicate the \xxx are stable under different number of selected vulnerable feature maps.

\paragraph{Perturbation.} In~\cref{algo:attack}, we will use $\epsilon$ to constrain the maximum perturbation of the FMA. To evaluate the impact of $\epsilon$ and $\alpha$ for the model, use four different $\epsilon$ and $\alpha$ setting to compare \xxx's effectivenss~(for ease of discussion, we set $\alpha = \epsilon/4$). The experiment results are shown in~\cref{tab:alpha}. We can observe that under different perturbation strength, \xxx's effectiveness do not have large change for each model. For instance, when $\alpha$ increase from $1/255$ to $4/255$, \xxx's FDR only change from 62.10\% to 62.29\% in MNIST\&LeNet1 combination, and the RAUC also only change from 87.97\% to 88.17\%, which indicate \xxx are stable under different perturbation size.

\mybox{Answer to RQ3: \xxx work stable under different hyper-parameter settings.}

\begin{table}[]
    \centering
    % \small
    \setlength{\tabcolsep}{1.7pt}
    \begin{tabular}{l|cccccccc}
        \toprule
        Dataset&\multicolumn{2}{c}{MNIST}& \multicolumn{2}{c}{Fashion}&\multicolumn{2}{c}{CIFAR10}&\multicolumn{2}{c}{SVHN} \\
        Model&LeNet1&LeNet5&LeNet1&ResNet20&VGG16&ResNet20&LeNet5&VGG16\\ 
         \midrule
        NAC &57.36&55.49&54.12&46.71&58.81&65.53&69.75&47.15\\
        KMNC&81.48&55.86&81.57&52.63&65.39&64.71&88.35&47.98\\
        NPC&50.68&61.29&52.78&63.15&62.07&65.07&65.61&49.84\\
        DSA&59.77&56.46&60.72&70.45&77.03&68.27&63.83&67.02\\
        Gini&85.00&85.33&82.96&62.28&62.23&78.53&91.54&\textbf{88.47}\\
        ATS&84.48&83.98&81.57&62.28&63.48&75.42&80.96&85.69\\
        RS  &74.01&65.37&70.04&\textbf{71.59}&74.15&72.66&90.02&74.62\\
        Our&\textbf{87.97} &\textbf{86.03} &\textbf{84.86} &70.07 &\textbf{77.25} &\textbf{79.42} &\textbf{95.17} & 86.11\\
        \bottomrule
    \end{tabular}
    \caption{When selecting 20\% test cases, the ratio of area under the curve~(RAUC) of FTCR plots.}
    % \vspace{-0.4cm}
    \label{tab:auc}
\end{table}

\begin{table}[]
% \small
\setlength{\tabcolsep}{1.7pt}
    \centering
    \begin{tabular}{c|cccc}
    \toprule
    &ADAPT&DeepM++&DeepH&Our\\
         Dataset\&Model&Cases/Acc(\%)&Cases/Acc(\%)&Cases/Acc(\%)&Cases/Acc(\%)\\
         \midrule
         MNIST~(LeNet1)&771/92.84&3071/92.88&2315/92.81&41196/93.45\\
         MNIST~(LeNet5)&2254/95.22&5032/96.31&2431/95.33&46048/97.92\\
         Fashion~(LeNet1)&6578/80.01&10571/82.77&7759/81.03&48202/85.36\\
         Fashion~(ResNet20)&1105/91.61&3086/91.83&1037/90.07&12422/91.75\\
         SVHN~(LeNet5)&11273/86.50&17533/86.37&13375/96.77&37724/87.59\\
         SVHN~(VGG16)&559/93.77&3989/93.20&1277/93.39&19411/94.21\\
         CIFAR10~(VGG16)&732/83.61&2554/84.33&1132/83.55&10244/86.96\\
         CIFAR10~(ResNet20)&1622/86.14&3488/86.93&2441/86.76&12255/88.86\\
         \bottomrule
    \end{tabular}
    \caption{\xxx and baselines generated fuzz test cases, which will be \textbf{misclassied} by DNN, within 5 minutes. In the repairing experiments, we randomly select the same number of cases as ADAPT to eliminate the effect of training case size~(e.g., we only select 771 cases from \xxx and other baseliens in MNIST \& LeNet1). DeepM++  and DeepH mean DeepMutation++ and DeepHyerion.}
    \label{tab:fuzz}
    % \vspace{-0.5cm}
\end{table}

\begin{table}[t]
    \centering
    % \small
    \setlength{\tabcolsep}{2pt}
    \begin{tabular}{l|c c c c c c}
    \toprule
         \multirow{2}{*}{Dataset~(DNN)}& \multicolumn{6}{c}{FDR of Top k vulnerable feature maps}\\
         &1&5&10&15&20&25\\
         \midrule
         MNIST~(LeNet-1)&56.15&62.10&62.72&62.30&62.10&62.01\\
         % MNIST~(LeNet-5)&51.45&52.8&52.5&52.15& \\
         % Fashion~(LeNet-1)& 56.8&57.7&58.2&-&-\\
         Fashion~(ResNet-20)&46.85&60.70&60.30&60.01&58.30&58.30 \\
         SVHN~(LeNet-5)&50.20&62.34&62.32&62.30&62.17&62.15\\
         % SVHN~(VGG-16)&\\
         % CIFAR-10~(VGG-16)\\
         CIFAR-10~(ResNet-20)&51.30&50.40&50.15&49.80&50.35&50.75\\
         \midrule
         \multirow{2}{*}{Dataset~(DNN)}& \multicolumn{6}{c}{RAUC of Top k vulnerable feature maps}\\
         &1&5&10&15&20&25\\
          \midrule
         MNIST~(LeNet-1)& 88.09&87.97&88.31&88.42&88.51&88.55\\
         % MNIST~(LeNet-5)& 86.03&88.05&88.59&88.84&\\
         % Fashion~(LeNet-1)& 84.86&83.76&83.97&-&-\\
         Fashion~(ResNet-20)&59.68&70.07&66.66&66.03&64.84&62.30\\
         SVHN~(LeNet-5)&84.78&95.17&94.82&95.23&94.46&95.10\\
         % SVHN~(VGG-16)&41.4\%&	43.1\%&	43.3\%&	45.7\% & 44.1\%\\
         % % CIFAR-10~(VGG-16)&55.7\%	&56.6\%&	56.9\%&	60.0\% & 58.2\%\\
         CIFAR-10~(ResNet-20)&76.89&78.78&81.11&78.89&81.63&85.93\\
         \bottomrule
    \end{tabular}
    \caption{FDR and FTCR with the different number of vulnerable feature maps ranging from 5 to 25, when 20\% of the test cases have been selected.
    We only report four combinations, while other combinations have similar results.}
    % \vspace{-0.5cm}
    \label{tab:size}
\end{table}

\begin{table}[t]
    \centering
    % \small
    % \setlength{\tabcolsep}{2pt}
    \begin{tabular}{l|c c  c c}
    \toprule
         \multirow{2}{*}{Dataset~(DNN)}& \multicolumn{4}{c}{FDR of different $\alpha$}\\
         &1/255&2/255&3/255&/4/255\\
         \midrule
         MNIST~(LeNet-1)&62.10&62.17&62.24&62.29\\

         Fashion~(ResNet-20)&60.70&60.78&60.85&60.90\\
         SVHN~(LeNet-5)&62.34&62.39&62.43&62.48\\
         % SVHN~(VGG-16)&\\
         % CIFAR-10~(VGG-16)\\
         CIFAR-10~(ResNet-20)&50.40&50.48&50.55&50.60\\
         \midrule
         \multirow{2}{*}{Dataset~(DNN)}& \multicolumn{4}{c}{RAUC of different $\alpha$}\\
         &1/255&2/255&3/255&/4/255\\
          \midrule
         MNIST~(LeNet-1)& 87.97&88.09&88.14&88.17\\
         % MNIST~(LeNet-5)& 86.03&88.05&88.59&88.84&\\
         % Fashion~(LeNet-1)& 84.86&83.76&83.97&-&-\\
         Fashion~(ResNet-20)&70.07&70.14&70.17&70.18\\
         SVHN~(LeNet-5)&95.17&95.30&95.41&94.44\\
         % SVHN~(VGG-16)&41.4\%&	43.1\%&	43.3\%&	45.7\% & 44.1\%\\
         % % CIFAR-10~(VGG-16)&55.7\%	&56.6\%&	56.9\%&	60.0\% & 58.2\%\\
         CIFAR-10~(ResNet-20)&78.78&78.89&78.89&78.90\\
         \bottomrule
    \end{tabular}
    \caption{FDR and FTCR with the different $\alpha$ and $\epsilon$.}
    % \vspace{-0.7cm}
    \label{tab:alpha}
\end{table}

\subsection{Threats to Validity}

\textbf{Test subject selection.} The selection of evaluation subjects (i.e., datasets and DNN models) could be a threat to validity. We try to counter this by using four commonly studied datasets (i.e., MNIST, Fashion, SVHN, and CIFAR10); for DNN models, we use four well-known pre-trained DNN-based models~(i.e., LeNet-1, LeNet-5, ResNet-20, and VGG-16) of different sizes and complexity ranging from 3,350 neurons up to more than 35,749,834 neurons. However, it doesn't guarantee that \xxx can be applied to all models. Additional models will be used to evaluate \xxx in future work.

\textbf{Data simulation.} Another threat to validity comes from the augmented test input generation. In order to generate test cases, we choose seven well-used benign mutations~(i.e., shift, rotation, scale, shear, contrast, brightness, and blur) as our baselines to simulate faults from different sources and granularity. Although these data simulations are very similar to virtual environment noise, it is impossible to guarantee that the distribution of the real unseen input is the same as our simulation. Additional experiments based on real unseen inputs need to be conducted in future work.

\textbf{Parameters settings.} The last threat could be the parameter settings in baselines. To compare with neuron coverage-guided and prioritization test set selection methods, we reproduced existing coverage methods for DNN, which may include user-defined parameters. By fine-tuning the parameter settings, the selected data could be different. To alleviate the potential bias, we follow the authors’ suggested settings or employ the default settings of the original papers. Due to page constrain, the impact of different $step\_size$ are not discussed in the paper, which will be evaluated in the future.

%% file: related.tex
\section{Related Work}
This section discusses the related work in two groups: testing the DNN model and deep learning testing.

\subsection{Neuron Coverage Metrics}
\citet{pei2017deepxplore} proposes the first white-box coverage criteria (i.e., NAC), which calculates the percentage of activated neurons. Since NAC is coase-grain neuron metric,
DeepGauge~\cite{ma2018deepgauge} then extends NAC and proposes a set of more fine-grained coverage criteria by considering the distribution of neuron outputs from the training data.
Noticing that DNNs have millions of neuron, testing all neuron is a large overhead, ~\cite{Gerasimou2020ImportanceDrivenDL} propose IDC, which focus on an important neuron in DNNs. Inspired by the coverage criteria in traditional software testing, some conventional coverage metrics~\cite{Deepcover,Ma2019DeepCTTC,Ma2018DeepMutationMT} are proposed. DeepCover~\cite{Deepcover} proposes the MC/DC coverage of DNNs based on the dependence between neurons in adjacent layers. To explore how inputs affect neuron internal decision logic flow,~\citet{Xie2022NPCNP} propose NPC to explore the neuron path coverage. DeepCT~\cite{Ma2019DeepCTTC} adopts the combinatorial testing idea and proposes a coverage metric that considers the combination of different neurons at each layer. DeepMutation~\cite{Ma2018DeepMutationMT} adopts the mutation testing into DL testing and proposes a set of operators to generate mutants of the DNN. Furthermore, DeepConcolic~\cite{Sun2018ConcolicTF} analyzed the limitation of existing coverage criteria and proposed a more fine-grained coverage metric that considers the relationships between two adjacent layers and the combinations of values of neurons at each layer.
Since neuron coverage testing can not detect fault-inducing feature maps in the DNN, in this work, we propose \xxx to address this problem, which can detect faults that are induced by the feature maps in the DNN~(\cref{sec:eval:selection}).

\subsection{Deep Learning Testing}
\citet{pei2017deepxplore} propose the first deep learning testing technique~(i.e., DeepXplore), which test DNN internal neuron activation parttern. Based on the DeepXplore's coverage metric, \citet{Tian2018DeepTestAT} propose DeepTest, which is used to generate test cases to explore DNN behaviors. Noticing DeepXplore is coarse-grained, \citet{ma2018deepgauge} propose DeepGuage, which tests DNN by fine-grained metrics. Based on the DeepGauge~\cite{ma2018deepgauge},~\citet{Ma2018DeepMutationMT} propose DeepMutation~\cite{Ma2018DeepMutationMT} and DeepMutation++~\cite{DeepMutation++}, these techniques are used to mutate test case to explore model incorrect behaviors. ~\citet{tensorfuzz} also uses metrics proposed by the above-mentioned coverage metric to generate test cases. Since the test cases generated are not always useful, and feeding all test cases into the model is time-consuming and the benign selection strategies used by coverage metrics are also time-consuming, 

Based on the above-mentioned testing techniques, some automated testing  techniques~\cite{Deepcover,tensorfuzz,pei2017deepxplore,Tian2018DeepTestAT,Sun2018ConcolicTF,DeepMutation++,Xie2018DeepHunterHD} are proposed to generate test inputs towards explore DNN incorrect behaviours. 
In addition, while the neuron coverage guided testing techniques are widely studied, the existing work~\cite{ismeanful,deepgini,li2019structural,gao2022adaptive} found that using NC-guided testing can not detect all types of faults in DNN. Such findings motivate this work that proposes a new type of test unit, i.e., feature map, to detect more types of faults that are not detected by neuron criteria testing techniques.

% DeepXplore~\cite{pei2017deepxplore} proposes the first white-box coverage criteria, i.e., Neuron Activation Coverage, which calculates the percentage of activated neurons. A differential testing approach is proposed to detect errors by increasing NAC. DeepGauge~\cite{ma2018deepgauge} then extends NAC and proposes a set of more fine-grained coverage criteria by considering the distribution of neuron outputs from the training data. Inspired by the coverage criteria in traditional software testing, some coverage metrics~\cite{Deepcover,Ma2019DeepCTTC,Ma2018DeepMutationMT} are proposed. DeepCover~\cite{Deepcover} proposes the MC/DC coverage of DNNs based on the dependence between neurons in adjacent layers. DeepCT~\cite{Ma2019DeepCTTC} adopts the combinatorial testing idea and proposes a coverage metric that considers the combination of different neurons at each layer. DeepMutation~\cite{Ma2018DeepMutationMT} adopts the mutation testing into DL testing and proposes a set of operators to generate mutants of the DNN. Furthermore, DeepConcolic~\cite{Sun2018ConcolicTF} analyzed the limitation of existing coverage criteria and proposed a more fine-grained coverage metric that considers both the relationships between two adjacent layers and the combinations of values of neurons at each layer. Compared to the aforementioned techniques that focus on exploring the neuron-level behaviors, the contribution of \xxx is that it proposes a novel test unit, i.e., feature maps, which could help DNN developer to explore more types of behaviors in DNNs.

%% file: conclusion.tex
\section{Conclusion}
In this work, we propose \xxx, which test DNNs from feature map level, for testing DNNs. Unlike existing neuron testing techniques, \xxx take the feature map as a testing unit, which delves into every inner feature maps that the models learned and detects vulnerable ones in the testing process.
The key component of \xxx is its testing metrics called \fvs and \fas. \fvs quantifies the vulnerability of the model feature maps, and \fas measures the impact of vulnerable feature maps on the model's accuracy. Then we propose a new feature map guided test case selection, which selects test cases by measuring each test case's \fvs value in vulnerable feature maps. 
Compared with coverage-guided and prioritization test case selection methods, \xxx's test case selection increases the fault detection rate by 49.97\% and 24.8\% on average, respectively.
Finally, we utilize the proposed \fvs metric to automatically fuzz for more valuable test cases to repair vulnerable feature maps and improve the model's accuracy.
% \section{Data Availability}
The source code of \xxx with all the evaluation details are available on our Github Page~\cite{DeepFeature}.